\documentclass{article} 
\usepackage[submission]{colm2025_conference}

\usepackage{tcolorbox}
\definecolor{chart}{HTML}{1f77b4}

\newtcolorbox{example}[1][]{
  colback=chart!5!white,
  colframe=chart,
  floatplacement=floating,
  title=\centering \textsf{#1}
}

\usepackage{adjustbox}
\usepackage{algorithm}
\usepackage{algpseudocode}
\usepackage{booktabs}
\usepackage{graphicx}
\usepackage{hyperref}
\usepackage{longtable}
\usepackage{microtype}
\usepackage{url}
\usepackage{lineno}
\usepackage{CJKutf8}
\usepackage{colortbl}
\usepackage{capt-of}
\usepackage{pifont}
\usepackage{enumitem}

\definecolor{darkblue}{rgb}{0, 0, 0.5}
\hypersetup{colorlinks=true, citecolor=darkblue, linkcolor=darkblue, urlcolor=darkblue}

\title{\bugs{}: Scaling Data for Software Engineering Agents}

\newcommand{\stanford}{\textcolor{red}{\textsuperscript{1}}}
\newcommand{\princeton}{\textcolor{orange}
{\textsuperscript{2}}}
\newcommand{\independent}{\textcolor{blue}{\textsuperscript{3}}}
\newcommand{\alibaba}{\textcolor{violet}{\textsuperscript{4}}}

\author{
John Yang\stanford,
Kilian Lieret\princeton,
Carlos E. Jimenez\princeton,
Alexander Wettig\princeton,
Kabir Khandpur\independent,
\And
\textbf{
Yanzhe Zhang\stanford,
Binyuan Hui\alibaba,
Ofir Press\princeton,
Ludwig Schmidt\stanford,
Diyi Yang\stanford
}
\And
\stanford Stanford University\quad
\princeton Princeton University\quad
\independent Indepedent\quad
\alibaba Alibaba Qwen
}

\newcommand{\bugs}{SWE-smith}
\newcommand{\greencheck}{\includegraphics[height=0.8em]{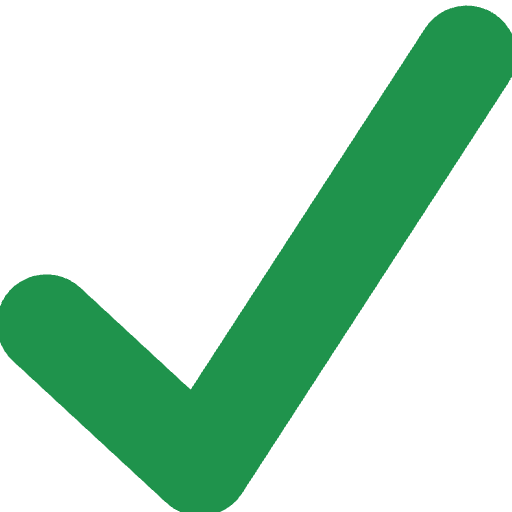}}
\newcommand{\redx}{\includegraphics[height=0.8em]{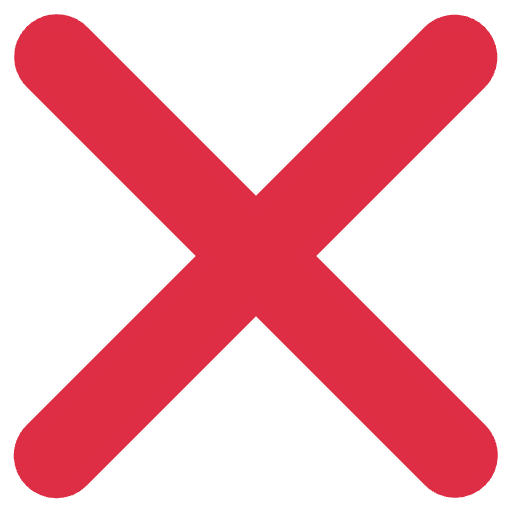}}

\newcommand*{\strreplaceview}{\texttt{str\_replace\_editor view}}

\colmfinaltrue
\begin{document}

\ifcolmsubmission
\fi

\maketitle

\begin{abstract}
Despite recent progress in Language Models (LMs) for software engineering, collecting training data remains a significant pain point.
Existing datasets are small, with at most $1{,}000$s of training instances from $11$ or fewer GitHub repositories.
The procedures to curate such datasets are often complex, necessitating hundreds of hours of human labor; companion execution environments also take up several terabytes of storage, severely limiting their scalability and usability.
To address this pain point, we introduce SWE-smith, a novel pipeline for generating software engineering training data at scale.
Given any Python codebase, \bugs{} constructs a corresponding execution environment, then automatically synthesizes $100$s to $1{,}000$s of task instances that break existing test(s) in the codebase.
Using \bugs{}, we create a dataset of $50$k instances sourced from $128$ GitHub repositories, an order of magnitude larger than all previous works.
We train \texttt{SWE-agent-LM-32B}, achieving $40.2$\% Pass@1 resolve rate on the SWE-bench Verified benchmark, state of the art among open source models.
We open source \bugs{} (collection procedure, task instances, trajectories, models) to lower the barrier of entry for research in LM systems for automated software engineering.
All assets available at \url{https://swesmith.com}.
\end{abstract}
\section{Introduction}

\begin{figure}[b]
\centering
\begin{minipage}[t]{0.49\textwidth}
  \centering
  \includegraphics[width=\linewidth]{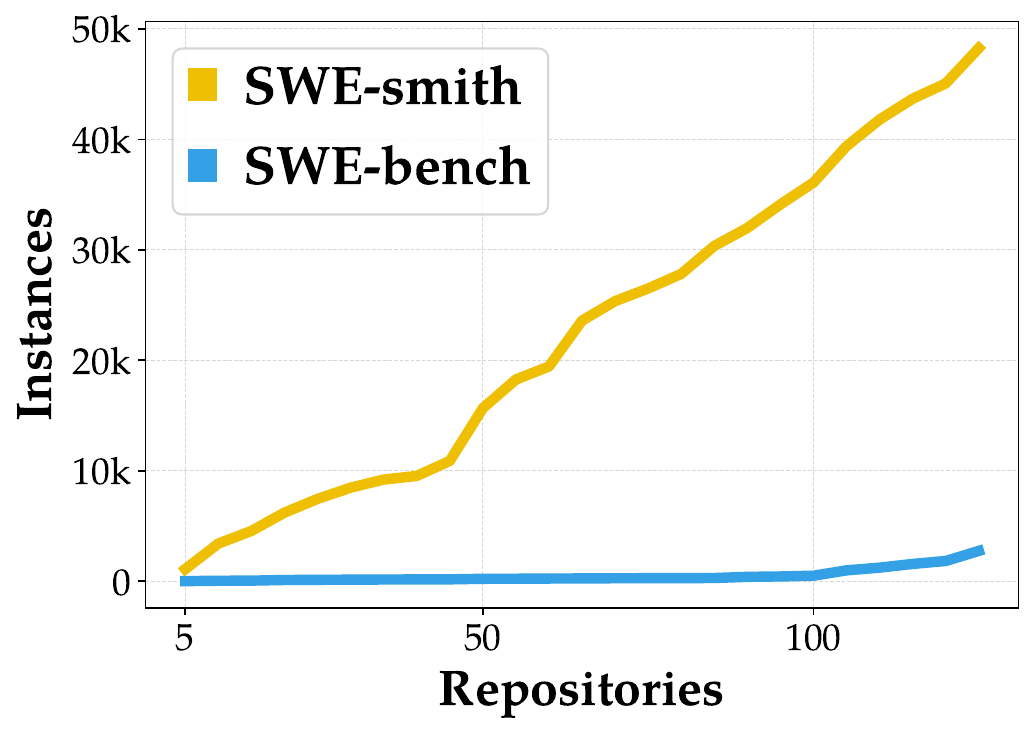}
\end{minipage}
\hfill
\begin{minipage}[t]{0.49\textwidth}
  \centering
  \includegraphics[width=\linewidth]{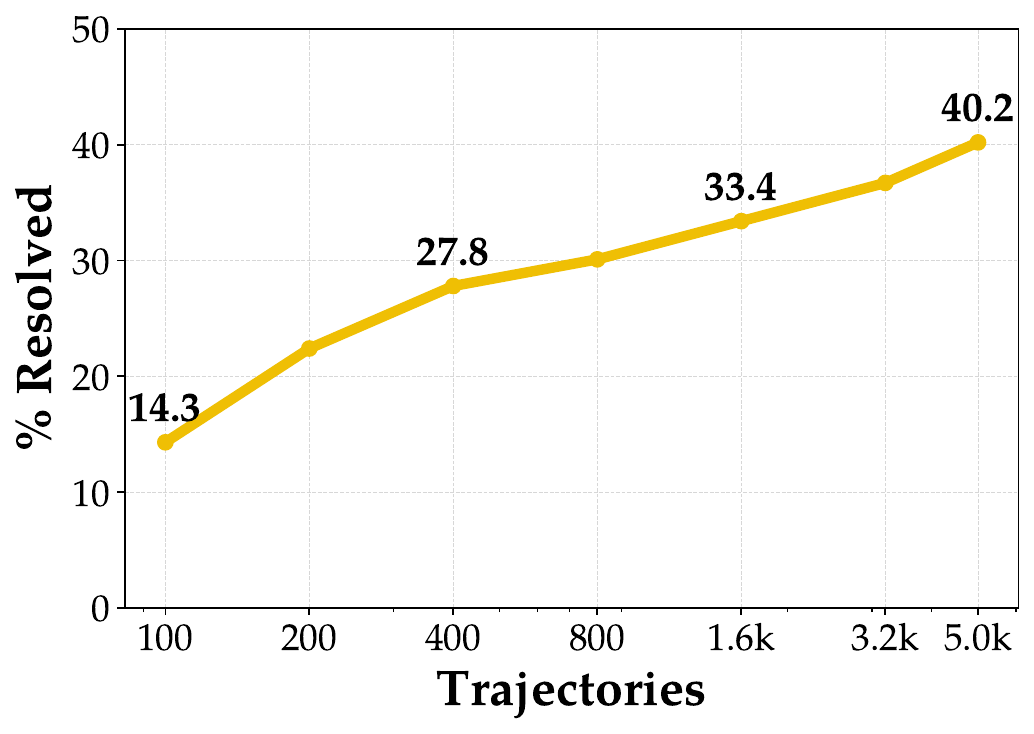}
\end{minipage}
\caption{
\textbf{Scaling task instances} (left) and \textbf{performance} (right) for SWE-agent's with \bugs{}.
Using \bugs{}, we can create $100$s to $1000$s of instances for any Python codebase, enabling us to train \texttt{SWE-agent-LM-32B} which achieves $40.2$\% on SWE-bench Verified.
}
\label{fig:scaling_teaser}
\end{figure}

Language Model (LM) agents, such as SWE-agent~\citep{yang_swe-agent_2024} or OpenHands~\citep{wang_openhands_2024}, 
have made remarkable progress towards automating software engineering (SE) tasks, as tracked by benchmarks such as SWE-bench~\citep{jimenez_swe-bench_2024}.
However, the most effective agents still rely on proprietary LMs, as building open source LMs for SE remains bottlenecked by the lack of large-scale, high-quality training data.
To ensure that open research remains relevant in this field, it is critical to develop infrastructure for collecting software engineering training data at scale.

The current open-source ecosystem offers two kinds of data sources to train LMs on SE tasks.
One simple approach is to crawl pull requests (PRs) and issues from GitHub repositories.
However, without execution environments or tests, these instances offer no reliable way of validating generated solutions,
and LMs are limited to learning from the surface form of code~\citep{xie2025swefixertrainingopensourcellms} or via rewards based on superficial string similarity~\citep{wei2025swerladvancingllmreasoning}.

In contrast, SWE-bench provides reliable validation by running unit tests against proposed solutions.
Another line of work has simply extended the SWE-bench collection strategy to a new set of repositories for training purposes \citep{pan_training_2024}. 
This produces flexible environments for training and distilling LM agents, since we can generate agent trajectories and filter them based on the unit test results.
However, the scalability of this approach is severely limited by the challenges associated with SWE-bench's collection strategy.
SWE-bench's filtering process leaves only a small number of PRs that not only resolve a Github issue, but also make meaningful changes to unit tests.
Also, setting up execution environments for each instance requires a substantial amount of human intervention.

In this paper, we introduce the \bugs{} toolkit, which marries the flexible execution environments of SWE-bench with scalable instance collection (Figure~\ref{fig:scaling_teaser}).
\bugs{} features several techniques to automatically synthesize bugs in existing GitHub repositories,
such as (1) generating errant rewrites of functions with an LM, (2) procedurally modifying the abstract syntax tree (AST) of functions, (3) undoing PRs, and (4) combining bugs.
Our key insight is that execution-based validation can not only validate proposed solutions,
but also identify bug candidates which cause substantial software regression (i.e., break tests).

\begin{figure}[t]
    \centering
    \includegraphics[width=\textwidth]{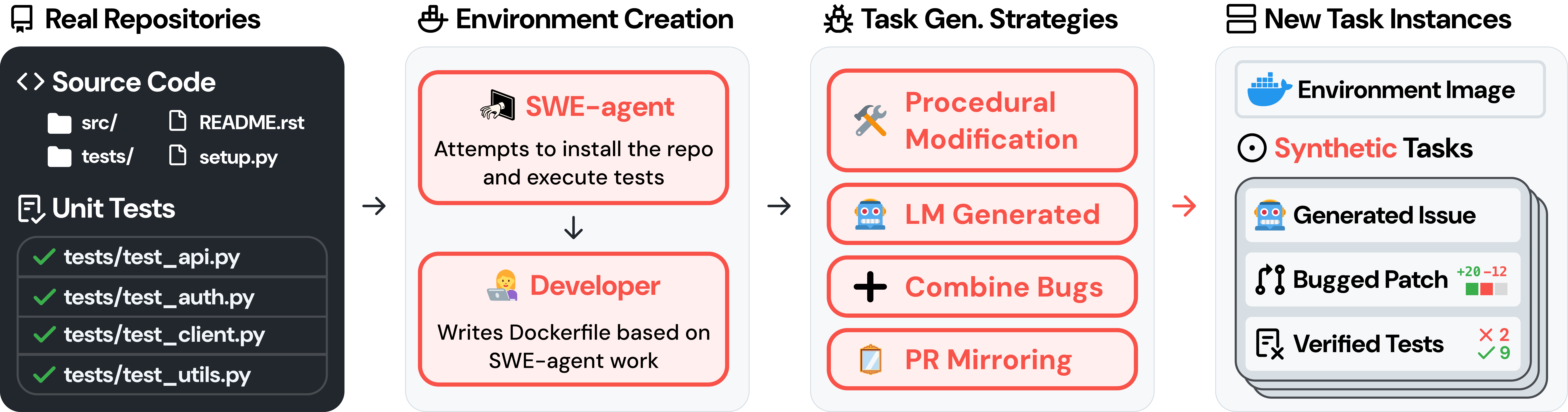}
    \caption{
    \bugs{} creates training data for software engineering agents by crafting bugs into real codebases.
    Given a codebase, we employ several strategies to create task instances that break existing tests.
    Using \bugs{}, we create $50$k+ task instances with execution environments from $128$ real world repositories.
    }
    \label{fig:preview}
\end{figure}

In a nutshell, \bugs{} puts forth the following task creation workflow, as shown in Figure~\ref{fig:preview}.
Given a codebase, we automatically set up a corresponding environment using SWE-agent~\citep{yang_swe-agent_2024}.
Within this environment, we then use the aforementioned techniques to synthesize $100$s to $1,000$s of task instances.
Finally, we craft realistic issue descriptions automatically with LMs.
\bugs{}'s design significantly reduces the amount of human labor and storage required for constructing execution environments.
Using \bugs{}, we create a dataset of $50$k task instances across $128$ real-world GitHub repositories.

Using the \bugs{} dataset, we achieve a new open-weight state of the art result on SWE-bench verified.
Using the SWE-smith task instances, we generate $5{,}016$ expert trajectories with Claude 3.7 Sonnet and fine-tune Qwen 2.5 Coder Instruct $32$B.
The resulting LM,  {\tt SWE-agent-LM-32B}, achieves $40.2\%$ (+$33.4$\%) on SWE-bench Verified in a single attempt, without inference-time scaling.
This sets a new state of the art for open-weight models.

The scale and diversity of the \bugs{} dataset enables us to begin establishing truths and investigate interesting phenomena about developing SWE-agents.
Training on more instances, bug types, and repositories helps.
LM generated issue text approximates real ones effectively.
Using \bugs{}, we find that it's possible to optimize LMs to perform well for specific repositories while only suffering minor generalization loss.

We release \bugs{} as an open-source toolkit --- including instances, environments, and trajectories --- to catalyze the development of stronger open-source LM agents.

\section{\bugs{}: Software Task Generation at Scale}
\label{sec:swesmith}
The core principle of \bugs{}'s collection strategy is to define an execution environment first, and then synthesize task instances within the environment.
Conceptually, this is a simple inversion of SWE-bench's approach, which instead prioritizes identifying task instances, and then attempts to build an environment for each.
In this section, we describe the procedure in detail and show how, in practice, \bugs{} scales significantly better in terms of repositories, task instances, and storage.

\subsection{Collection}
\label{sec:swesmith:collection}

\textbf{Building execution environments for repositories with passing tests.}
Given a repository, we run SWE-agent~\citep{yang_swe-agent_2024} on the latest commit for at most $100$ steps, instructing it to install the codebase and run the test suite.
We then manually verify the installation and testing instructions, check if more than $80$\% of existing tests pass, and finally create a Docker image for the repository.
We target repositories for the $5,000$ most downloaded packages listed in the Python Package Index (PyPI) as of November 18, 2024, sort the PyPI packages by GitHub stars, and then remove any PyPI package with less than $1,000$ stars, as well as all $12$ SWE-bench test repositories from consideration.
More in \S\ref{appx:infra:repo_selection}.

\textbf{Creating task instance candidates.}
Per repository, we employ four different strategies to create candidates.
As shown in Figure~\ref{fig:preview}, each strategy takes in a repository as input, then produces task instance candidates represented as \texttt{.diff} files.
Extensive details in \S\ref{appx:bugs:generate}.
\begin{itemize}[leftmargin=15pt]
    \item \textbf{LM Generation}: Per repository, we identify all programmatic entities (functions, classes), then take two approaches: (1) provide an LM with the function and prompt it to introduce errant \textit{modifications} (henceforth referred to as ``LM Modify"), and (2) given only the function header and docstring, ask the LM to \textit{rewrite} it (``LM Rewrite").
    More in \S\ref{appx:bugs:generate:lm}.
    \item \textbf{Procedural Modification}: Per function, we acquire an abstract syntax tree (AST) representation of the code, then randomly perform one or more transformations (e.g., remove a conditional/loop, change an operator, +$11$ more. See Table~\ref{tab:pm_list}).
    More in \S\ref{appx:bugs:generate:prod}.
    \item \textbf{Combine Bugs}: LM generation and Procedural Modification task instances exclusively edit one function or class.
    To create more complex tasks that require editing multiple portions of the codebase, we devise a ``Patch Combination" strategy that creates a task instance by aggregating candidates from the same file(s) or module(s).
    More in \S\ref{appx:bugs:generate:combine}.
    \item \textbf{Invert PRs} (or ``PR Mirror"): Per repository, we collect all PRs that modify Python files.
    Per PR, we attempt to \emph{undo} its revisions in the current version of the repository.
    To achieve this, we provide an LM with the PR's code changes (a \texttt{.diff} plaintext) and prompt it to rewrite each affected file such that the PR edits are reverted.
    Unlike SWE-bench, we do \textit{not} check out the PR’s base commit, as the install specifications determined in the previous step may not be compatible with older versions of the repo.
    More in \S\ref{appx:bugs:generate:pr}.
\end{itemize}

\textbf{Execution-based validation of candidates.}
We apply each candidate patch to the corresponding repository, run the test suite, and only keep patches that break one or more existing, passing tests (referred to as \textit{Fail-to-Pass} or \textit{F2P} test(s)).
For efficiency purposes, we also limit testing runtime to two minutes; bug candidates that cause test runtimes in excess of this time limit are discarded.
Minor additional details in \S\ref{appx:infra:harnesses}.

\textbf{Generating problem statements.} 
The issue text associated with a bug can significantly alter the difficulty and feasibility of the task instance.
Detailed descriptions of ``expected" vs. ``observed" behavior or bug-reproduction code in issue text heavily affect an agent's capacity to localize bugs or iterate on proposed solutions.
We explore several techniques covered fully in \S\ref{appx:issue_generation}, and ultimately settle on a simple strategy.
Per task instance, we provide an LM with the \texttt{.diff} patch, source code of a random F2P test, and execution output from running the repository's test suite with the bug patch applied.
We prompt the LM for GitHub issue-style text that includes reproduction code based on the F2P test.

\textbf{What human labor remains?}
The steps requiring manual effort are (1) parsing the correct installation setup procedures from the agent trajectory ($\sim7$ min per repository), 
and (2) implementing the parser for test outputs ($\sim1$ min per repository).
Step two requires very little time because parsers can be reused for repositories with the same testing infrastructure (e.g., \texttt{pytest}).
\bugs{} removes the need for manual efforts to determine installation specifications for multiple versions of a codebase across time, the most costly step of SWE-bench collection.
Creating \bugs{} took one author $\sim{}20$h of human labor.

\subsection{Features}
\label{sec:swesmith:features}

We apply \bugs{} to $128$ Python repositories, generating a total of $50$k instances.
Table~\ref{tab:dataset_summary} captures the key statistics.
On average, we generate $381$ task instances per repository, with as many as $2277$ for \texttt{pandas-dev/pandas}.
We summarize the distribution of task instances per repository in Figure~\ref{fig:num_insts_per_category}, where repositories are grouped into one of six general categories.
\bugs{} took \$$1360$ to create (\$$1000$ to generate bugs, \$$160$ for automatic repository installation with SWE-agent, \$$200$ to generate issues for $10$K bugs).
Generating an issue costs $2.54$¢ on average.
More dataset analyses in \S\ref{appx:dataset}.

\begin{table}[t]
\centering
\begin{minipage}[b]{0.33\textwidth}
    \setlength{\abovecaptionskip}{0em}
    \includegraphics[width=\textwidth]{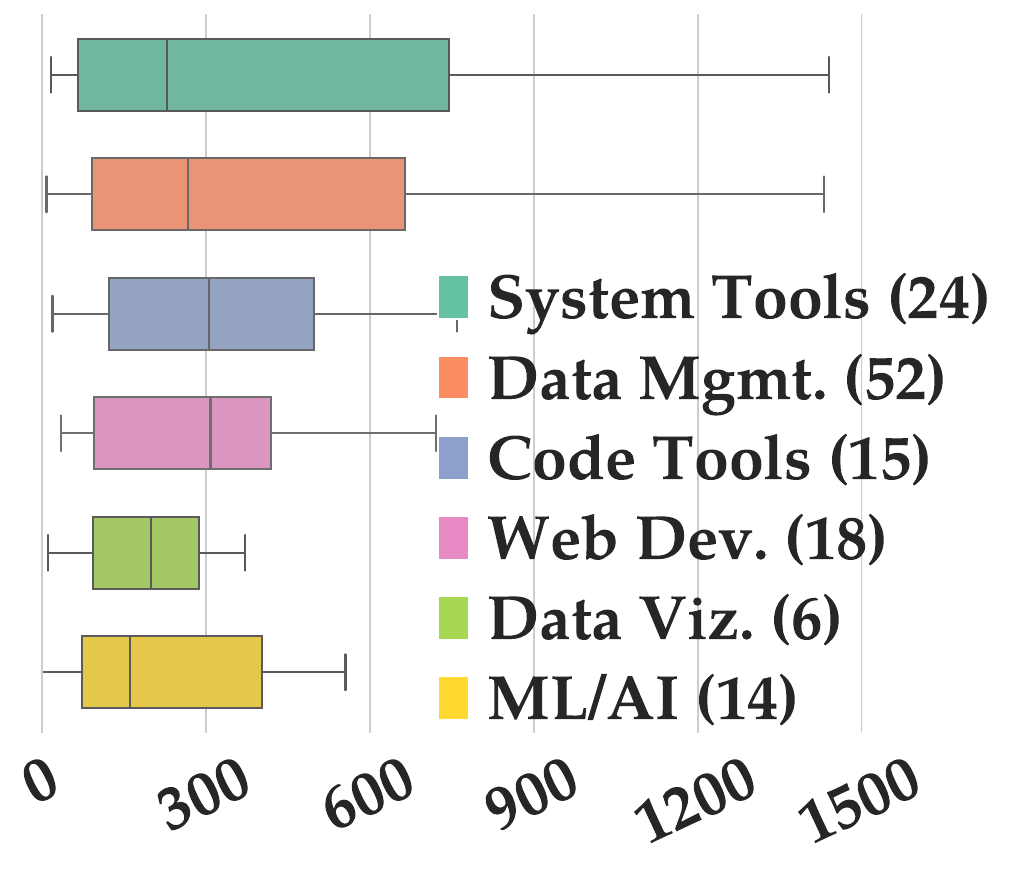}
    \captionof{figure}{
    Distribution of instances per repo for $128$ repo's grouped into $6$ categories.
    }
    \label{fig:num_insts_per_category}
\end{minipage}
\hfill
\begin{minipage}[b]{0.64\textwidth}
    \centering
    \begin{tabular}{l|ll|rrr}
\toprule
Bug Type & Yield \% & \# Insts & Cost & F2P & Lines \\
\midrule
Combine    & $96.9$\% & $10,092$ & $0.00$¢ & $15$ & $11$ \\
LM Modify  & $56.0$\% & $17,887$ & $0.38$¢ & $4$  & $3$ \\
LM Rewrite & $35.0$\% & $4,173$  & $3.93$¢ & $4$  & $24$ \\
PR Mirror  & $33.8$\% & $2,344$  & $5.53$¢ & $3$  & $14$ \\
Procedural & $40.2$\% & $15,641$ & $0.00$¢ & $7$  & $5$ \\
\midrule
Total & 50.1 & 50,137 & 2.32¢ & 6 & 5 \\
\bottomrule
    \end{tabular}
    \caption{
    Summary of \bugs{} statistics.
    ``Yield \%" is the \% of candidates generated by a strategy that break $1+$ tests.
    ``Cost" is the average cost to generate one candidate.
    ``F2P" (Fail to Pass tests), ``Lines [Edited]" are median values.
    }
    \label{tab:dataset_summary}

\end{minipage}
\end{table}

Bug generation strategies vary in cost and yield rate.
Of methods relying on LMs, PR Mirrors are more expensive because the task entails rewriting entire files, as opposed to individual functions for LM Modify and LM Rewrite.
Yield rates are limited by either lack of test coverage for the change or because the bug candidate did not actually introduce relevant issues.
For example, for LM Rewrite, the LM is asked to re-implement the function; it is \textit{not} explicitly asked for bugs.
When requested outright (LM Modify), the yield is higher.

\begin{table}[b]
    \centering
    \begin{tabular}{l|ccccc}
\toprule
Dataset  & \# Tasks & \# Repos &  Exec? & Source & Env. Size \\
\midrule
R2E \scriptsize~\citep{jain2024r2e} & $0.25$k  & $137$ & \greencheck & Synth & $270$ GBs \\
R2E-gym (Subset) \scriptsize~\citep{jain2025r2e-gym} & $4.6$k & $10$  & \greencheck & Synth & $4$ TBs \\
SWE-bench-extra \scriptsize~\citep{badertdinov2024scaling} & $6.38$k & $2$k & \redx & Real & - \\
SWE-bench-train \scriptsize ~\citep{jimenez_swe-bench_2024} & $19$k  & $37$  & \redx & Real & - \\
SWE-fixer \scriptsize~\citep{xie2025swefixertrainingopensourcellms} & $115$k & $856$ & \redx & Real & - \\
SWE-gym \scriptsize~\citep{pan_training_2024} & $2.4$k & $11$ & \greencheck & Real & $6$ TBs\\
\midrule
\textbf{\bugs{}} & $50$k & $128$ & \greencheck & Both & $295$ GBs\\
\bottomrule
    \end{tabular}
    \caption{
Comparison of open source training datasets for software engineering tasks.
Relative to existing datasets, \bugs{} has multiple times the number of task instances, repositories, and environments at a fraction of prior storage costs.
SWE-fixer and SWE-bench-train task instances do not have execution environments, so ``Env. Size" is blank.
    }
    \label{tab:dataset_comparison}
\end{table}

\textbf{How difficult are \bugs{} task instances?}
To determine whether task instances produced by \bugs{} are realistic and challenging, we train a Qwen $2.5$ $32$B model on $1{,}699$ human-annotated (task, label) pairs from~\citet{chowdhuryintroducing} to rate tasks as (\texttt{easy}, \texttt{medium}, \texttt{hard}) by training.
To quantify difficulty, each difficulty label corresponds to values of $1$/$5$/$9$.
The model achieves $75.3$\% test accuracy.
We then rate difficulty of task instances from both \bugs{} and prior SWE-bench style datasets~\citep{chowdhuryintroducing,jimenez_swe-bench_2024,pan_training_2024,yang_swe-bench_2024}.
\bugs{} task instances span a broad range of difficulties, similar to SWE-bench and SWE-gym.
The average difficulty score for \bugs{} ($5.27$–$5.72$ across bug generation strategies) is comparable to SWE-bench ($5.01$) and SWE-gym ($5.62$). This suggests SWE-smith enables realistic and appropriately challenging evaluation.
We discuss why bug strategies yield different levels of difficulty and visualize difficulty per dataset in \S\ref{appx:difficulty}.

\textbf{Scaling execution environments.}
Unlike SWE-bench which creates a Docker image per task instance, \bugs{} leverages a simpler design where tasks from the same repository share the same environment, reducing storage overhead significantly, as shown in Table~\ref{tab:dataset_comparison}.
This approach not only makes scaling task instances more affordable, but also renders \bugs{} more accessible and maintainable than existing datasets.
We estimate that creating a similar quantity of task instances ($50$k) using SWE-bench would require $50$ to $150$ TBs of storage for environments, a $500$x difference.
Extended discussion in \S\ref{appx:dataset:bug_gen_stats}.

\section{Experiments}
\label{sec:experiments}
To explore the utility of \bugs{} for training software engineering agents, we use rejection sampling fine-tuning~\citep{yuan2023scalingrelationshiplearningmathematical} as the primary procedure for improving a base LM with \bugs{}.
Our experiment workflow is as follows.
First, we curate a subset of \bugs{} task instances.
Next, we run an agent system with an expert model on this subset.
At this step, the trajectory corresponding to each run is recorded.
Then, we fine-tune the base (or ``student") model on the trajectories corresponding to resolved instances.
Finally, we evaluate the agent system run with the student model on a separate, test split.

\textbf{Models.} For expert models, we use \texttt{claude-3-7-sonnet-20250219}~\citep{anthropicclaude37}.
For fair comparisons with prior works~\citep{pan_training_2024}, we also use \texttt{claude-3-5-sonnet-20240620} and \texttt{gpt-4o-2024-08-06}.
We use the \texttt{Qwen-2.5-Coder-Instruct}~\citep{hui2024qwen25codertechnicalreport} $7$B and $32$B series as the base models.
Training and hyperparameter details are in \S\ref{appx:experiments:train_details}.

\textbf{Agent system.} We use SWE-agent~\citep{yang_swe-agent_2024}, an agent system for solving GitHub issues.
SWE-agent provides a base LM with an Agent Computer Interface (ACI) that enables more effective interactions with a codebase.
At each turn, SWE-agent prompts an LM to generate a ReAct~\citep{yao_react_2023} style (thought, action) pair, where the action either edits a file or executes a shell command.
We choose SWE-agent because, at the time of writing, SWE-agent with Claude 3.7 Sonnet is the top open source solution on SWE-bench.
When generating trajectories with expert models, we run SWE-agent for at most $75$ steps and \$$2.00$ cost limit.
For inference of student models, we impose the same $75$ step maximum and fix temperature at $0.0$.
Full configuration details are in \S\ref{appx:experiments:train_details}.

\textbf{Evaluation metrics.}
We evaluate on the SWE-bench Lite and Verified~\citep{chowdhuryintroducing} subsets.
SWE-bench evaluates AI systems on their ability to solve software issues from $12$ real world GitHub repositories.
The Lite split is a subset of $300$ instances, curated to be an easier evaluation set that's less costly to run.
The Verified split is a human-curated subset of $500$ instances, selected for clearer problem statements and more reliable evaluation.
To assess generalization beyond Python, we also evaluate on SWE-bench Multilingual, a new dataset introduced in this paper.
SWE-Bench Multilingual consists of $300$ task instances that cover $9$ additional programming languages.
See \S\ref{appx:experiments:eval_datasets} for more details.
We report the \textbf{\% resolved} metric, the proportion of successfully resolved instances.
\section{Results}
\label{sec:results}
Table~\ref{tab:main_results} compares the performance of Qwen~2.5 Coder Instruct models (7B and 32B), fine-tuned on $5{,}016$ \bugs{} trajectories.
We refer to them as {\tt SWE-agent-LM-7B} and {\tt SWE-agent-LM-32B}; the latter achieves state-of-the-art performance.

\begin{table}[ht]
    \centering
    \begin{tabular}{ll|ccc}
\toprule
Model & System & Train Size & Lite & Verified  \\
\midrule
\multicolumn{5}{c}{\textit{Closed Weight Models}} \\
\midrule
GPT-4o~\citep{openai2024gpt4ocard} & Agentless & - & $32.0$ & $38.8$ \\
                    & OpenHands & - & $22.0$ & -      \\
                    & SWE-agent & - & $18.3$ & $23.0$ \\
Claude 3.5 Sonnet~\citep{anthropicclaude35} & Agentless & - & $40.7$ & $50.8$ \\
                    & AutoCodeRover & - & -  & $46.2$ \\
                    & OpenHands & - & $41.7$ & $53.0$ \\
                    & SWE-agent & - & $23.0$ & $33.6$ \\
Claude 3.7 Sonnet~\citep{anthropicclaude37}   & SWE-agent & - & \textbf{48.0} & \textbf{58.2} \\
Llama3-SWE-RL-70B~\citep{wei2025swerladvancingllmreasoning}  & Agentless & 11M & -      & $41.0$ \\
\midrule
\multicolumn{5}{c}{\textit{Open Weight Models}} \\
\midrule
Lingma-SWE-GPT-72B~\citep{ma2024lingmaswegptopendevelopmentprocesscentric} & SWE-SynInfer & - & - & $28.8$ \\
Qwen3-235B-A22B~\citep{qwen_qwen25_2025} & OpenHands & - & - & $34.4$ \\
R2E-Gym-32B~\citep{jain2025r2e-gym} & OpenHands & $3.3$k & -      & $34.4$ \\
SWE-fixer-72B~\citep{xie2025swefixertrainingopensourcellms} & SWE-Fixer & $110$k & $24.7$ & $32.8$ \\
SWE-gym-32B~\citep{pan_training_2024} & OpenHands & $491$  & $15.3$ & $20.6$ \\
SWE-agent-LM-7B     & SWE-agent & $2$k   & $11.7$ & $15.2$  \\
SWE-agent-LM-32B    & SWE-agent & $5$k   & \textbf{30.7} & \textbf{40.2} \\
\bottomrule
    \end{tabular}
    \caption{
    Resolve rates for existing solutions on SWE-bench Lite and Verified, collected from ~\citet{jimenez2024leaderboard}, compared to models fine-tuned on \bugs{}.
    All performance numbers are pass@$1$.
    We do \textit{not} compare against systems that use verifiers or multiple attempts at test time.
    }
    \label{tab:main_results}
\end{table}

The final dataset of $5{,}016$ training points was curated as follows.
We start by collecting a large pool of expert trajectories.
First, we carried out each of the ablations in Section~\ref{sec:results:ablations}, giving us an initial set of $5{,}105$ trajectories.
Next, based on our observation that PR Mirror and LM Rewrite task instances yield the most effective expert trajectories (discussed below), we run the expert model on all task instances of these types, bumping up the total number to $6{,}457$ task instances.
Ultimately, we attempt to generate expert trajectories for $8{,}686$ unique task instances, or $17.3$\% of the \bugs{} dataset.
Reinforcing the difficulty rating findings from Section~\ref{sec:swesmith:features}, we observe that \bugs{} task instances are non-trivial for the top agent systems today.
The final pool of $6{,}457$ represents a $36$\% resolve rate of all $17{,}906$ attempts to solve one of the $8{,}686$ task instances.

Next, we perform minor filtering of this collection.
As reported in~\citet{pan_training_2024}, we also observe that ``easier" trajectories -- task instances that are repeatedly solved across multiple runs --- degrade model performance.
Therefore, we limit the number of times any \bugs{} task instance is represented in the training set to $3$ trajectories.
This leads to the final $5{,}016$ training set.
More details in \S\ref{appx:experiments:data_breakdown}.

\textbf{Performance improves with more data points.}
Extending similar graphs from ~\citet{jain2025r2e-gym,pan_training_2024}, Figure~\ref{fig:scaling_teaser} shows increasing performance with more trajectories.


\textbf{Comparison at the same training set size.}
To compare with prior works~\citep{jain2025r2e-gym,pan_training_2024}, we run expert trajectory generation on $1000$ random \bugs{} task instances with SWE-agent + Claude 3.5 Sonnet ($800$) or GPT-4o ($200$).
We then fine-tune the $32$B model on $500$ successful trajectories, a training set size both works report on.
Our model achieves a $28.2$\% resolve rate on SWE-bench Verified, a relative difference of $+8.2$\% with \citet{pan_training_2024} and $+0.7$\% with \citet{jain2025r2e-gym}.

\subsection{Ablations of \bugs{}}
\label{sec:results:ablations}
We perform several ablations of how \bugs{}'s bug and problem statement generation strategies impact the quality of training data.
We use Claude 3.7 Sonnet as the expert for fine-tuning Qwen $2.5$ $7$B Coder Instruct, and report the performance on SWE-bench Verified.

\textbf{LM Rewrite and Procedural bugs are comparable to PR mirrors.}
We randomly sample $1000$ instances per bug generation strategy (LM Modify, LM Rewrite, Procedural Modifications, PR Mirrors).
Per instance, we generate issue text with an LM and run expert trajectory generation.
We then fine-tune a student model per strategy, capping training points to the minimum number of successful trajectories from any strategy ($507$) for fair comparison.

Table~\ref{tab:ablation_bug} summarizes the results.
Trajectories generated from PR mirrors are empirically the most effective training data --- this is expected, since they are most reflective of SWE-bench.
What's noteworthy is that trajectories from Procedural Modification and LM Rewrite instances lead to competitive models.
There is a steep drop-off with LM Modify bugs.

\textbf{LM generated issues are comparable to real issues.}
We randomly sample $600$ PR Mirror task instances.
We compare LM generated issues with three alternatives --- fixed issue templates, the source code + test logs of a random Fail-to-Pass test, and the original issue text associated with the PR.
We again cap training points to the minimum number of successful trajectories ($259$) for fairness.

As shown in Table~\ref{tab:ablation_issue}, training on task instances with LM generated issues is empirically comparable to using the original issue text.
Using fixed issue templates not only leads to the fewest successful trajectories, but also results in relatively homogeneous problem solving sequences.
The expert trajectories from fixed issue templates have $31$\% fewer unique actions compared to LM generated text ($379$ vs. $550$).
While providing a Fail-to-Pass test case leads to more successful expert trajectories, leaking the evaluation criteria causes the model to skip over writing a reproduction script, which accounts for the performance drop.
Of $500$ SWE-bench Verified instances, the student model trained on LM-generated issues attempts to reproduce the bug for $379$ of the runs.
The model trained on test-based issues only does so for $127$ cases, a $66$\% decrease.

\begin{table}[t]
\centering
\begin{minipage}[t]{0.52\textwidth}
    \centering
    \begin{tabular}{l|cc}
\toprule
Strategy & \# Trajs. & \% Resolved \\
\midrule
LM Modify  & $802$ &  $5.7$ ($\pm1.5$) \\
LM Rewrite & $507$ &  $8.8$ ($\pm1.7$) \\
Procedural & $745$ &  $8.6$ ($\pm1.8$) \\
PR Mirror  & $557$ &  $9.2$ ($\pm1.7)$ \\
\bottomrule
    \end{tabular}
    \caption{
Comparison of training on $1000$ \bugs{} instances created with different strategies.
    }
    \label{tab:ablation_bug}

\end{minipage}
\hfill
\begin{minipage}[t]{0.45\textwidth}
    \centering
    \begin{tabular}{l|cc}
\toprule
Issue & \# Trajs. & \% Resolved \\
\midrule
Fixed     & 259 & $6.4$ ($\pm 1.5$) \\
F2P Test  & 390 & $7.3$ ($\pm 1.9$) \\
LM        & 328 & $7.7$ ($\pm 1.5$) \\
Original  & 319 & $7.8$ ($\pm 1.8$) \\
\bottomrule
\end{tabular}
\caption{
Comparing training on $600$ PR Mirror instances with varied issue text.
}
    \label{tab:ablation_issue}
\end{minipage}
\end{table}

\textbf{Task difficulty correlates with solvability but not with effectiveness as training data.}
First, we run our difficulty rating model on $10$k randomly selected \bugs{} task instances.
From this pool, we curate subsets of $1000$ instances corresponding to the three difficulty levels, then run expert trajectory generation per subset $3$ times.
For the \texttt{easy}/\texttt{medium}/\texttt{hard} subsets, the resolve rate by the expert model are $58.6$\%, $41.0$\%, and $17.0$\% respectively.

Next, from all successful trajectories, we create four fine-tuning datasets of $500$ trajectories each corresponding to difficulty scores of $2$, $4$, $6$, and $8$.
As mentioned in Section~\ref{sec:swesmith:features}, the corresponding scores for \texttt{easy}/\texttt{medium}/\texttt{hard} are $1$/$5$/$9$.
Therefore, the SFT dataset for score $2$ is made up of trajectories corresponding to $375$ \texttt{easy} and $125$ \texttt{medium} instances, and so on.
Somewhat surprisingly, we do not observe strong correlation between increased difficulty and downstream performance.
For the student models trained on the $2$/$4$/$6$/$8$ difficulty SFT datasets, we get pass@1 scores of $12.4$\%, $10.8$\%, $13.6$\%, and $12.2$\% on SWE-bench Verified.

\begin{figure}[b]
    \centering
    \begin{minipage}[t]{0.56\textwidth}
        \centering
        \includegraphics[width=\textwidth]{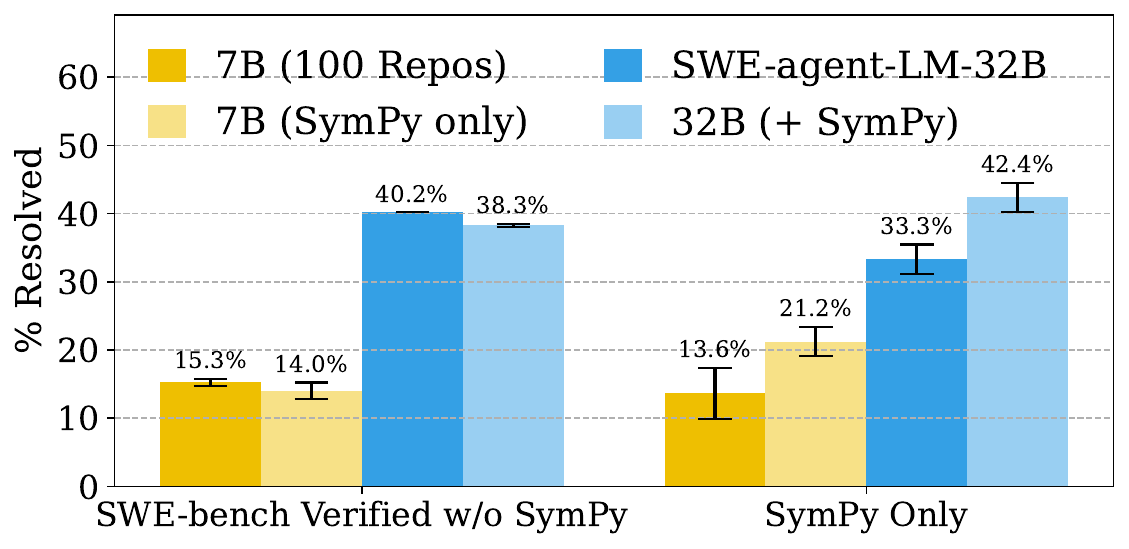}
        \caption{
        We fine-tune a $7$B base and our $32$B models on $700$ trajectories for SymPy.
        Specialization boosts performance with minor generalization loss.
        }
        \vspace{0.3em}
        \label{fig:specialized_ablation}
    \end{minipage}%
    \hfill
    \begin{minipage}[t]{0.42\textwidth}
        \centering
        \includegraphics[width=\textwidth]{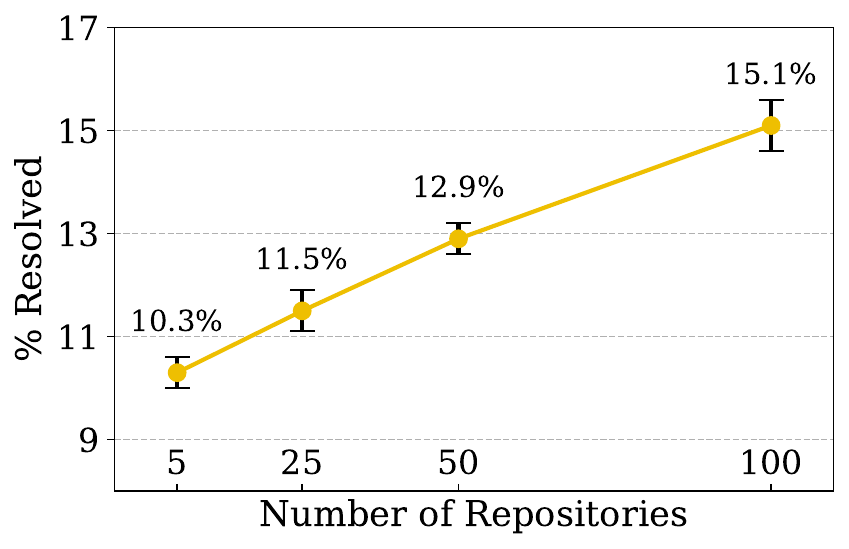}
        \caption{
        At $700$ training samples, we observe performance increases logarithmically with repositories.
        }
        \label{fig:repo_diversity_ablation}
    \end{minipage}
\end{figure}

\textbf{Training on more repositories improves general performance.}
We train models in four settings by sampling $700$ expert trajectories on Procedural Modification tasks from pools of $4$, $25$, $50$, and $100$ repositories.
Echoing similar findings for code generation tasks~\citep{xie2025repostscalablerepositorylevelcoding}, we find that increasing repositories represented in the training set improves performance, as shown in Figure~\ref{fig:repo_diversity_ablation}, with an approximately logarithmic relation between model performance and number of repositories.

\textbf{Repository-specialized models excel on the target repository with minor generalization loss.}
We experiment with training models to be specialists on one particular repository.
To assess performance, we evaluate models on a subset of SWE-bench Verified tasks that are (1) from SymPy, and (2) created after January 1st, 2022, a total of $22$ instances.
To create SymPy specific training data, we first select a base commit of SymPy just before the cutoff date.
Next, we create $1276$ Procedural Modification task instances, then generate $700$ expert trajectories.
We evaluate specialization in two settings: (1) single-repository fine-tuning, and (2) specialist stage fine-tuning, both shown in Figure~\ref{fig:specialized_ablation}.
For single-repository tuning, we compare a model initialized with \texttt{Qwen-2.5-Coder-Instruct 7B} and trained on $700$ instances sampled from $100$ repositories, to the same Qwen base model but fine-tuned on the $700$ SymPy instances only.
For specialist stage fine-tuning, we simply compare \texttt{SWE-agent-LM-32B} to the same model further fine-tuned on the $700$ SymPy instances.

Specialization significantly boosts performance for the target repository with only slight drops in general performance in both the single-repository fine-tuning ($21.2$\% vs. $13.6$\%) and specialist stage fine-tuning ($42.4$\% vs. $33.3$\%) settings.

\subsection{Analysis of Agent Behavior}
\label{sec:results:agent_behavior}

\begin{figure}[t]
    \centering
    \begin{minipage}[t]{0.49\textwidth}
        \centering
        {\includegraphics[scale=0.48]{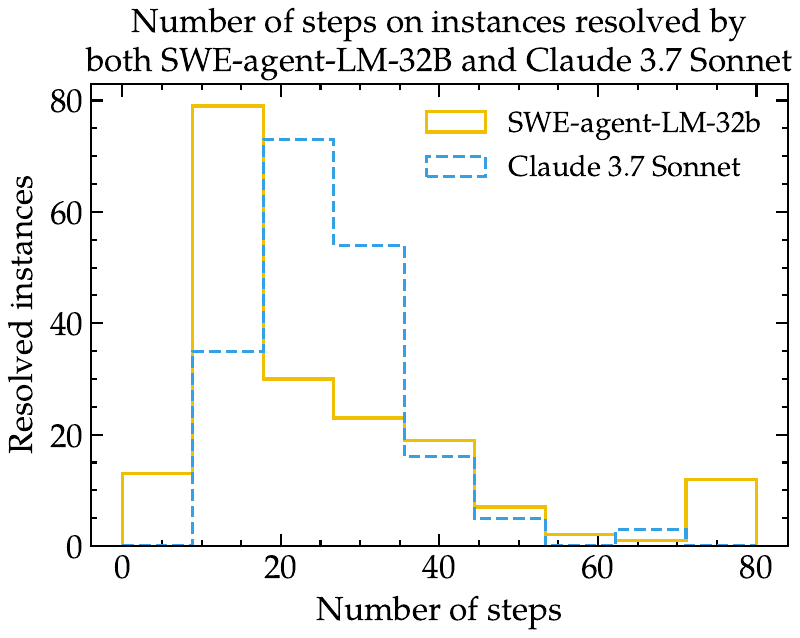}}
        \caption{
        \texttt{SWE-agent-LM-32B} takes fewer steps to submit compared to Claude 3.7 Sonnet for instances resolved by both models.
        }
        \label{fig:step_counts_overlap}
    \end{minipage}
    \hfill
    \begin{minipage}[t]{0.48\textwidth}
        \centering
        {\includegraphics[scale=0.48]{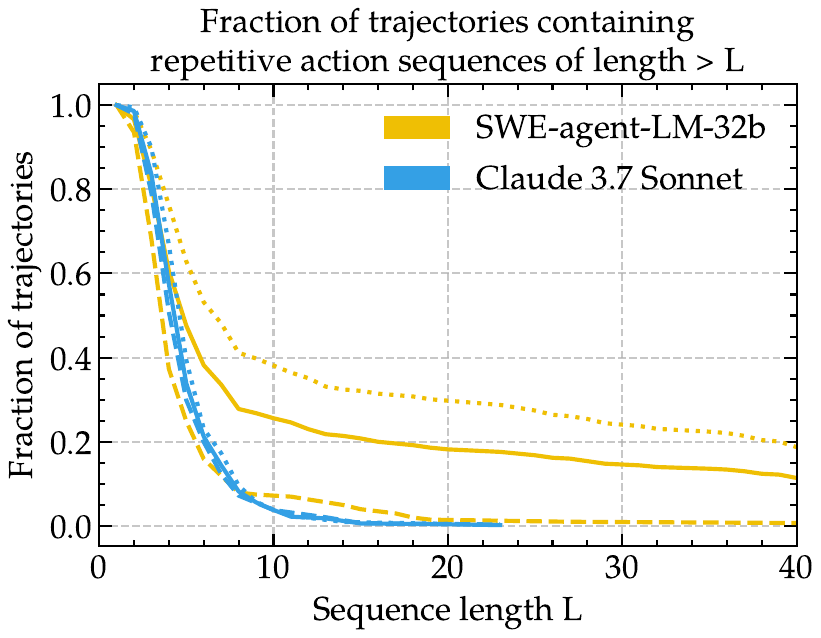}}
        \caption{
For unsuccessfully resolved tasks, a frequent failure mode is that \texttt{SWE-agent-LM-32B} will repeat actions.
        }
        \label{fig:repetitive_actions}
    \end{minipage}
\end{figure}

This section analyzes the behavior, failure modes, and efficiency of SWE-agent when run with \texttt{SWE-agent-LM-32B} or Claude 3.7 Sonnet on SWE-bench verified.

\textbf{SWE-agent-LM-32B can solve tasks efficiently.} \texttt{SWE-agent-LM-32B} resolves tasks in fewer steps on average (24.9) than Claude 3.7 Sonnet (29.1), though the difference becomes marginal when accounting for different average difficulties of the resolved tasks: On the overlap of tasks that are resolved by both LMs, \texttt{SWE-agent-LM-32B} uses 24.8 steps compared to 25.6 used by Claude 3.7 Sonnet (see Fig.~\ref{fig:step_counts_overlap}).
While shorter trajectories are not always preferred (additional actions can be used for additional validation purposes, for example), this shows that \texttt{SWE-agent-LM-32B} solves tasks very efficiently.
At the same time \texttt{SWE-agent-LM-32B} also demonstrates that it can remain focused throughout long trajectories, with 31 instances being resolved after 40 steps or more.  
We further highlight that the accuracy of naturally terminating
\footnote{i.e., excluding agent runs that are terminated due to errors or cost/step count limits. Note that SWE-agent still extracts and submits any changes performed by the agent in these cases and some of them can be successful (for example if the agent is terminated due to cost while testing already performed edits).}
agent submissions with \texttt{SWE-agent-LM-32B} achieve an accuracy nearly matching that of Claude 3.7 Sonnet (60\% vs 63\%), showing that \texttt{SWE-agent-LM-32B} is adept at determining whether an instance has been resolved.
As the overall cost and turn count averages scale strongly with the cost and turn limits, we reserve a more thorough analysis for \S\ref{appendix:experiments:agentbehavior:turncounts}.

\textbf{Repetitive actions are a key problem.}
We observe a tendency for \texttt{SWE-agent-LM-32B} to get stuck in long sequences of repetitive actions, in particular long sequences of calls that display different portions of a file instead of using search commands.
\footnote{In fact, these \strreplaceview{} commands make up $73$\% of the longest repetitive sequences. For this analysis, we look at repetitions of the base command, i.e., without any arguments. See \S\ref{appendix:experiments:agentbehavior} for more.}
More than 25\% of \texttt{SWE-agent-LM-32B} trajectories have a repetitive sequence of at least length 10, compared to less than 4\% for Claude 3.7 Sonnet (see Figure~\ref{fig:repetitive_actions}). 
The occurrence of long repetitive sequences correlates strongly with the agent's ability to solve the corresponding task instance, largely because the LM continues issuing similar commands until either the agent cost or turn limit is reached, at which point the run is terminated.
For example, repetitive sequences of length 10 correspond to an 89\% failure probability.
Simple interventions from the agent scaffold can mitigate repetitive actions, but do not seem to improve resolve rates (see~\S\ref{appendix:experiments:agentbehavior}).

\begin{figure}[t]
\centering
    \includegraphics[width=\textwidth]{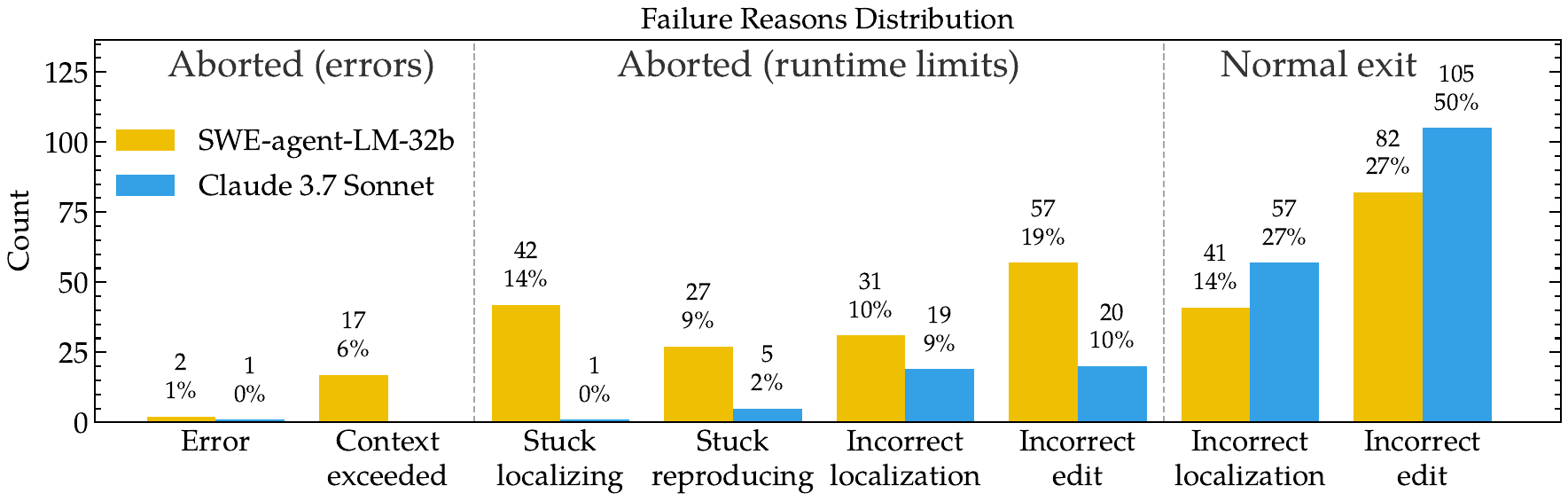}
    \caption{
    More than half of the unresolved instances of \texttt{SWE-agent-LM-32B} correspond to runs terminated by cost/step limits, and these limits are frequently reached before source code has been modified. See \S\ref{appendix:experiments:agentbehavior} for more.
    }
    \label{fig:failure_modes}
\end{figure}

\textbf{Localization is the dominant failure mode.}
Guided by a short plan in the system prompt, SWE-agent typically starts by \emph{localizing} (search and read actions), \emph{reproducing} (test file creation and execution), before modifying source files and validating the fixes.
If the agent gets stuck at any of these stages or keeps on iterating, the agent loop is eventually interrupted by runtime limits (cost, number of LM calls, runtime).
While this rarely happens with Claude 3.7 Sonnet, 53\% of \texttt{SWE-agent-LM-32b}'s failures are associated with such limits (Figure~\ref{fig:failure_modes}).
The agent often already gets stuck during localization or initial efforts to reproduce a bug, with endlessly repeated actions being a persistent issue.
More on failure modes in \S\ref{appendix:experiments:agentbehavior}. 


\section{Related Work}
\label{sec:related}
\textbf{LMs for Software Engineering.} As contemporary LMs have saturated traditional code generation tasks~\citep{austin_program_2021,chen_evaluating_2021}, software engineering benchmarks~\citep{jain2024r2e,jimenez_swe-bench_2024,yang_swe-bench_2024,zhao_commit0_2024,zan2025multiswebenchmultilingualbenchmarkissue}, notably SWE-bench, have become a new de facto evaluation setting due to their diverse, complex, real-world programming challenges.
The most significant source of open source progress on SWE-bench has been the development of LM-based workflows~\citep{orwall2024moatless,xia_agentless_2024,zhang_autocoderover_2024} and agents~\citep{antoniades_swe-search_2024,wang_openhands_2024,yang_swe-agent_2024,zhang_diversity_2024}.
Workflow-based systems are typically human-engineered decompositions of a task into a sequence of sub-goals.
\citet{yang_swe-bench_2024} suggests such pipelines may not generalize effectively to non-Python repositories, requiring additional human intervention to re-adapt.
We therefore elect to focus on generating trajectories with and for LM agent systems~\citep{sumers2024cognitivearchitectureslanguageagents,yang_intercode_2023,yao_react_2023}.
Because no workflow is imposed, agent systems inherently rely more on the LM to plan and refine its actions, putting more focus on an LM's capabilities, not inference scaffolds.

\textbf{Training Datasets for Coding.}
Prior work around training data has focused on instruction following ~\citep{luo2023wizardcoderempoweringcodelarge,muennighoff_octopack_2024,shypula2024learningperformanceimprovingcodeedits,wei_selfcodealign_2024,wei2024magicoderempoweringcodegeneration,yu2024wavecoderwidespreadversatileenhancement} and preference learning~\citep{liu2024learningcodepreferencesynthetic,liu_dstc_2024} for code completion tasks.
Several recent works introduce training sets for retrieval augmented generation~\citep{jimenez_swe-bench_2024,xie2025swefixertrainingopensourcellms}, workflows~\citep{wei2025swerladvancingllmreasoning}, and agent~\citep{badertdinov2024scaling,ma2024lingmaswegptopendevelopmentprocesscentric,pan_training_2024,jain2025r2e-gym} approaches to SWE-bench.
Our work applies~\citet{haluptzok2023languagemodelsteachprogram} at a repository level: by having an LM break a codebase, we drastically reduce the human effort needed to define a task and build its environment.
Concurrent to our work, \citet{xie2025repostscalablerepositorylevelcoding} (RePOST) also constructs execution environments for repository functions, but differs significantly in methodology and evaluation.
RePOST sandboxes a function and its dependencies to a separate script, then generates tests with an LM, removing the original codebase as context.
The tasks' source is repository-level; the environments and tasks are not.
RePOST evaluates solely on code generation (e.g., HumanEval~\citep{chen_evaluating_2021}).
\citet{jain2025r2e-gym} (R2E-Gym) improves open source LMs' performance on SWE-bench with inference time scaling and verifiers.
R2E-gym's $51$\% resolve rate is not comparable to Table~\ref{tab:main_results} results, as each instance is attempted $26$ times.
R2E-gym's $4.6$k training instances are collected using SWE-bench's pipeline, with some augmentations around using LMs to synthesize issue text and tests.
To our knowledge, we are the first to address the limited scalability
of previous approaches.

\section{Discussion}
\label{sec:discussion}

\textbf{Limitations and future directions.}
First, \bugs{}'s collection pipeline is Python-centric.
The mechanisms to identify programmatic objects (e.g. functions, classes) and perform transformations rely heavily on the Python specific \texttt{ast} library.
That said, \bugs{}'s collection strategy is transferable to other languages.
Second, due to both compute/budget constraints and our work's primary stance as a dataset contribution, we only include fine-tuning as a demonstration of \bugs{}'s effectiveness.
We do not explore other training techniques such as reasoning capabilities elicited via reinforcement learning.

\textbf{Conclusion.}
We introduce \bugs{}, a dataset of $50$k software engineering task instances from across $128$ real world GitHub repositories.
\bugs{} collection pipeline allows us to scale up task instances, environments, and trajectories at a fraction of prior costs without sacrificing faithfulness to open source software development practices.
Using \bugs{}, we train \texttt{SWE-agent-LM-32B}, achieving a state-of-the-art $40.2$\% on SWE-bench Verified.
Our experiments show how \bugs{} can be used to identify fundamental trends about developing SWE-agents.
We believe \bugs{} provides the foundational data and infrastructure needed to train software engineering agents in a truly scalable manner.

\section*{Acknowledgments}
We thank Princeton Language \& Intelligence (PLI) for providing credits for running closed-source API models.
Thanks to Samuel Ainsworth for his constant support of \texttt{bitbop.io} (\url{https://bitbop.io/}), the compute service for which the majority of the project was carried out with.
We'd also like to thank Akshat Bubna, Howard Halim, Andrew Liu, Peyton Walters, and the great team at Modal (\url{https://modal.com/}) for providing credits that made fine-tuning and model serving efforts extremely easy for this project.
This work is partially supported by ONR grant N000142412532 and NSF grant IIS-2247357.
We also thank Open Philanthropy and Andreessen Horowitz for providing funding for this work.
Finally, thanks to Tianyu Gao, William Held, Niklas Muennighoff, Rafael Rafailov, Yijia Shao, Chenglei Si, Anikait Singh, Tianyi Zhang, Kexin Pei, and Karthik Narasimhan for constructive discussions and support throughout this project.

\bibliography{colm2025_conference}
\bibliographystyle{colm2025_conference}

\newpage
\appendix
\section*{Appendix}
The appendix is generally structured as follows.
In Sections~\ref{appx:infra} to~\ref{appx:issue_generation}, we review details about \bugs{}'s infrastructure and collection strategies for curating the \bugs{} task instances and execution environments, providing comparisons to existing datasets such as SWE-bench and SWE-gym along the way.
In Sections~\ref{appx:difficulty} and onward, we discuss more about how we created the trajectories dataset, then provide additional ablations and results showcasing the effectiveness of \bugs{} as a dataset.

\begin{figure}[h]
    \centering
    \includegraphics[width=\textwidth]{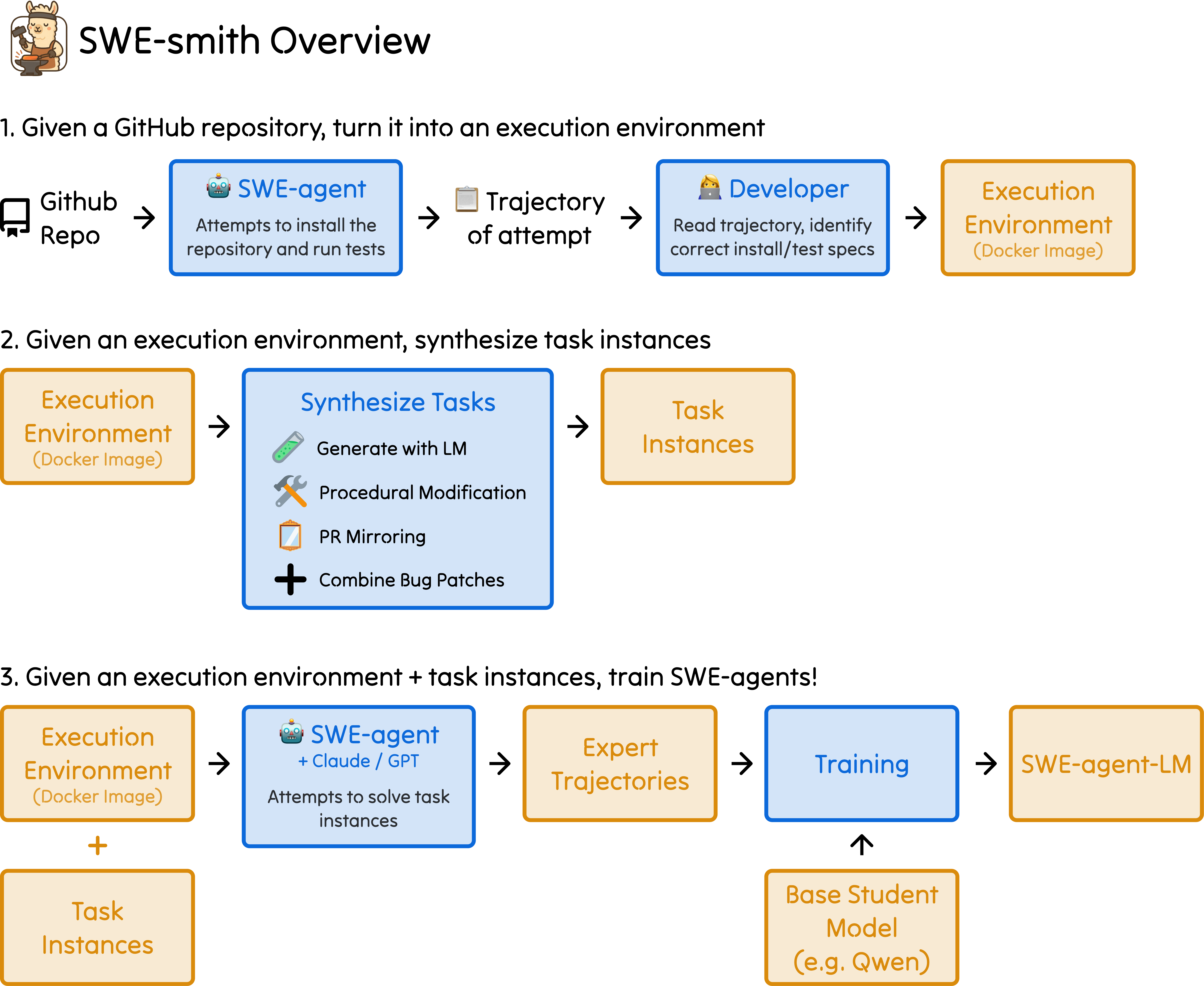}
    \caption{
    An overview of pipelines in \bugs{}.
    Scripts/functions and manual steps are highlighted in \textcolor{blue}{blue}.
    Artifacts that are also the inputs and outputs of these scripts are in \textcolor{orange}{orange}.
    \bugs{} fits in seamlessly with the SWE-bench and SWE-agent ecosystem.
    Use \bugs{} to construct execution environments and generate task instances.
    Use SWE-agent to generate expert trajectories on \bugs{} task instances and run inference with models trained on these trajectories.
    Use SWE-bench to evaluate how good your models are at resolving GitHub issues and performing software engineering tasks.
    }
    \label{fig:overview}
\end{figure}

\section{Infrastructure}
\label{appx:infra}
We cover additional details about how \bugs{} works, specifically
\begin{itemize}
    \item The form factor of a \bugs{} task instance.
    \item How we identify repositories and the SWE-agent configuration we use to automatically install them.
    \item How the task validation and evaluation harnesses work.
\end{itemize}

\subsection{\bugs{} Task Instance}
\label{appx:infra:anatomy}
We briefly review the format of a \bugs{} task instance, highlight how it is different from a SWE-bench task instance, and discuss why \bugs{}'s relatively simple infrastructure compared to SWE-bench allows us scale task collection much more efficiently.

A \bugs{} task instance is very similar to the form factor of a SWE-bench task instance, with several minor differences.
A \bugs{} task instance includes the following fields:
\begin{itemize}
    \item \texttt{repo}: The repository the task instance is from.
    \item \texttt{instance\_id}: A unique identifier (usually \texttt{(repo).(bug\_type).(hash)})
    \item \texttt{base\_commit}: Hash of the GitHub branch that points to the repository with the bug \texttt{patch} applied.
    \item \texttt{patch}: The \texttt{diff} that causes the bug. It is applied to the original codebase to create the bug. Reverting this patch is effectively the solution.
    \item \texttt{problem\_statement}: The generated issue text that conveys the bug.
    It is provided to a model or system before it begins attempting a fix.
    \item \texttt{created\_at}: A timestamp matching when the bug was successfully validated and pushed to the mirror repository as a branch.
    \item \texttt{FAIL\_TO\_PASS}: The unit tests that break when the test suite is run with the bug \texttt{patch} applied.
    \item \texttt{PASS\_TO\_PASS}: The unit tests that do not break.
    These correspond to the set of all tests minus the \texttt{FAIL\_TO\_PASS} tests.
\end{itemize}

We summarize the key distinctions between a \bugs{} and SWE-bench task instance:
\begin{itemize}
    \item \bugs{} task instances do not include the \texttt{version} or \texttt{environment\_setup\_commit} fields, which SWE-bench requires as additional identifiers for specifying repository-specific installation instructions across time.
    In \bugs{}, unique installation instructions are specified for each (repository, commit).
    \item The \texttt{hints\_text} field is not included. In SWE-bench, this refers to the issue and PR thread comments written after the first commit of the corresponding PR.
    \item The \texttt{created\_at} field is assigned the timestamp reflecting when the bug was successfully validated.
    Originally, \texttt{created\_at} refers to when a PR was created.
    \item There is no \texttt{test\_patch} field, as the \bugs{} collection pipeline does not create or synthesize any hidden tests.
    All \texttt{FAIL\_TO\_PASS} bugs are visible and runnable in the repository at inference time.
\end{itemize}

\subsection{Repository Selection}
\label{appx:infra:repo_selection}
In addition to the criteria discussed in Section~\ref{sec:swesmith:collection}, we also ensure that a repository has a license that allows non-proprietary use.
The majority of software licenses are permissive (BSD, MIT, Apache), while the remainder are largely protective licenses (GPL) that still allow for non-commercial use.
We inspected the repositories with custom licenses and confirmed they allowed for the use cases exercised in our work.
The licenses for each repository are fully listed in Table~\ref{tab:licenses}.

\begin{table}[ht!]
    \centering
    \resizebox{\textwidth}{!}{%
\newcolumntype{P}[1]{>{\raggedright\arraybackslash}p{#1}} 
\begin{tabular}{P{0.15\textwidth} | P{0.85\textwidth}}

\toprule
Apache License 2.0 & \footnotesize\texttt{Project-MONAI/MONAI; alanjds/drf-nested-routers; arrow-py/arrow; buriy/python-readability; facebookresearch/fvcore; getmoto/moto; google/textfsm; iterative/dvc; jax-ml/jax; jd/tenacity; kayak/pypika; modin-project/modin; pyca/pyopenssl; spulec/freezegun; tkrajina/gpxpy; tornadoweb/tornado; weaveworks/grafanalib} \\
\midrule
BSD 2-Clause "Simplified" License & \footnotesize\texttt{madzak/python-json-logger; pyasn1/pyasn1; pygments/pygments; sunpy/sunpy} \\
\midrule
BSD 3-Clause "New" or "Revised" License & \footnotesize\texttt{Suor/funcy; alecthomas/voluptuous; andialbrecht/sqlparse; cookiecutter/cookiecutter; dask/dask; django/channels; django/daphne; encode/starlette; gawel/pyquery; gweis/isodate; john-kurkowski/tldextract; lepture/mistune; oauthlib/oauthlib; pallets/click; pallets/flask; pallets/jinja; pallets/markupsafe; pandas-dev/pandas; scrapy/scrapy; theskumar/python-dotenv} \\
\midrule
GNU General Public License v3.0 & \footnotesize\texttt{Cog-Creators/Red-DiscordBot; adrienverge/yamllint} \\
\midrule
GNU Lesser General Public License v2.1 & \footnotesize\texttt{chardet/chardet; paramiko/paramiko; pylint-dev/astroid} \\
\midrule
GNU Lesser General Public License v3.0 & \footnotesize\texttt{Knio/dominate} \\
\midrule
ISC License & \footnotesize\texttt{kennethreitz/records} \\
\midrule
MIT License & \footnotesize\texttt{amueller/word\_cloud; borntyping/python-colorlog; bottlepy/bottle; cantools/cantools; cdgriffith/Box; cknd/stackprinter; conan-io/conan; cool-RR/PySnooper; datamade/usaddress; dbader/schedule; erikrose/parsimonious; facebookresearch/hydra; facelessuser/soupsieve; getnikola/nikola; graphql-python/graphene; hukkin/tomli; jaraco/inflect; jawah/charset\_normalizer; joke2k/faker; keleshev/schema; life4/textdistance; luozhouyang/python-string-similarity; marshmallow-code/apispec; marshmallow-code/marshmallow; marshmallow-code/webargs; martinblech/xmltodict; matthewwithanm/python-markdownify; mewwts/addict; mido/mido; mozillazg/python-pinyin; msiemens/tinydb; pdfminer/pdfminer; pndurette/gTTS; pudo/dataset; pydantic/pydantic; pyparsing/pyparsing; pytest-dev/iniconfig; python-hyper/h11; python-jsonschema/jsonschema; python-openxml/python-docx; pyupio/safety; pyvista/pyvista; r1chardj0n3s/parse; rsalmei/alive-progress; rubik/radon; rustedpy/result; scanny/python-pptx; seatgeek/thefuzz; sloria/environs; sqlfluff/sqlfluff; termcolor/termcolor; tobymao/sqlglot; tox-dev/pipdeptree; tweepy/tweepy; un33k/python-slugify; vi3k6i5/flashtext} \\
\midrule
Other & \footnotesize\texttt{Mimino666/langdetect; PyCQA/flake8; agronholm/exceptiongroup; agronholm/typeguard; aio-libs/async-timeout; benoitc/gunicorn; cloudpipe/cloudpickle; davidhalter/parso; django-money/django-money; gruns/furl; kurtmckee/feedparser; lincolnloop/python-qrcode; mahmoud/boltons; mahmoud/glom; mozilla/bleach; pexpect/ptyprocess; prettytable/prettytable; pwaller/pyfiglet; pydata/patsy; pydicom/pydicom; python-trio/trio; python/mypy; pyutils/line\_profiler; seperman/deepdiff} \\
\bottomrule
    \end{tabular}
    }
    \caption{License associated with each repository as of April 8, 2025. All licenses are permissive and allow for public, nonprofit use.}
    \label{tab:licenses}
\end{table}

We deliberately limit the search scope for repositories to those predominantly written in Python.
Following precedents, focusing on Python repositories allowed us to form assumptions about installation and testing procedures (e.g. repository is organized as a PyPI package, \texttt{pytest} is the testing framework) that made scaling up automatic repository setup with SWE-agent more tractable.
A worthwhile direction to consider for future work is expanding the coverage of repositories to be more comprehensive of codebases written in different programming languages, as ~\citet{yang_swe-bench_2024} does, extending SWE-bench style evaluation to JavaScript repositories with multimodal inputs.

\textbf{Automated repository installation.}
The goal of this step is to first, get the installation and testing instructions for a repository, and second, create a Docker image containing the repository with the development environment set up.

We provide the system prompt given to SWE-agent that asks it to install a repository in Figure~\ref{fig:prompt_install_repo}.
Each repository installation task is initialized with a clone of the original repository.
No additional steps (e.g. \texttt{pypi} package downloads, \texttt{conda} environment setup) are performed.

We run SWE-agent with \texttt{claude-3-5-sonnet-20241022} with a maximum cost limit of \$$2$ and a maximum call limit of $150$.
The installation run terminates whenever one of these conditions is met.
For every run, we record the interactions.
We then manually review the trajectory, identifying the appropriate installation and testing specifications.

Each run incurs an average cost of \$$0.72$ and an average of $17$ steps before SWE-agent issues the \texttt{submit} command.
The runs typically finish within two minutes.
The majority of Python repositories require fewer steps --- typically, SWE-agent will view the \texttt{CONTRIBUTING.md}, run the installation command provided verbatim in the text, and then runs \texttt{pytest}, showing all tests passing.
A minority of repositories will require several steps because additional dependencies must be installed with \texttt{apt-get}.
The manual review process following this requires $3$ to $20$ minutes.
One author carried out this effort for $128$ repositories, taking an estimated $18$ human hours to accomplish.
In the process of reaching $128$ repositories, the author gave up on $17$ repositories at the manual review stage.

\begin{example}[
System prompt for generating bugs with an LM
]
\small
\texttt{$<$uploaded\_files$>$} \\
\texttt{\{\{working\_dir\}\}} \\
\texttt{$<$/uploaded\_files$>$} \\
I've uploaded a python code repository in the directory \texttt{\{\{working\_dir\}\}}. \\

Can you please install this repository?
Your goal should be to configure the repository's development environment such that existing tests pass.
You are currently in the root directory of the repository, and nothing has been installed yet.
You in an Ubuntu 22.04 environment. \\

The repository is predominantly written in Python. Here are several tips for installing it: \\

1. A good place to start is to look for a \texttt{CONTRIBUTING.[md$\vert$rst]} file, which will often contain instructions on how to install the repository and any dependencies it may have. Occasionally, the \texttt{README.md} file may also contain installation instructions. \\

2. Usually, a repository may have \texttt{setup.py} or \texttt{pyproject.toml} files which can be used to install the package. \texttt{pip install -e .} is commonly used, although many packages will also require an additional specifier that installs development packages as well (e.g. \texttt{pip install -e .[dev]}). \\

3. To check whether the repository was installed successfully, run tests and see if they pass. You can usually find tests in a \texttt{tests/} or \texttt{test/} directory. You can run tests using \texttt{pytest} or \texttt{unittest}, depending on the framework used by the repository. \\

4. Sometimes, you will need to install additional packages, often listed in a \texttt{requirements.txt} or \texttt{environment.yml} file. Also, be mindful of Ubuntu system dependencies that may need to be installed via \texttt{apt-get} (e.g. \texttt{sudo apt-get install $<$package$>$}). \\

Once you are finished with installing the repository, run the \texttt{submit} command to submit your changes for review.
\end{example}

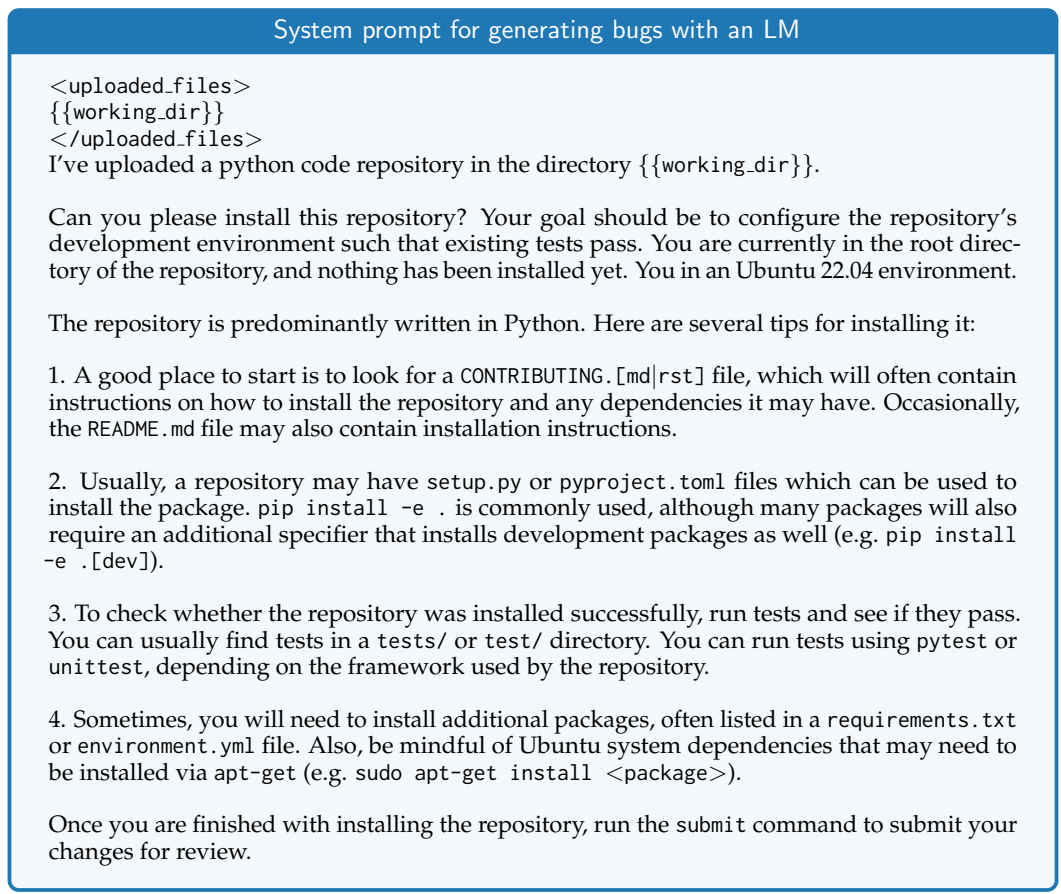
\captionof{figure}{
Prompt provided to SWE-agent + an LM asking it to install a repository.
}
\label{fig:prompt_install_repo}

\subsection{Validation, Evaluation Harnesses}
\label{appx:infra:harnesses}
We adapt SWE-bench's validation script to convert each bug patch into a SWE-bench style task instance.
This step ensures \bugs{} can be run by existing SWE-bench solutions.
The conversion involves two steps.
First, the bug patch is applied and pushed as a branch to a mirror clone of the repository.
Second, we create a SWE-bench style task instance from the bug patch, populating important fields such as Fail-to-Pass and Pass-to-Pass tests with information from the validation logs.

\newpage
\section{Bug Generation Strategies}
\label{appx:bugs:generate}
In this section, we review each of the bug generation strategies we employ in depth.
While we experimented with several bug generation strategies, the ones we elect to include are those we found to satisfy several desirable properties.

\begin{enumerate}
    \item The approach works in a codebase-agnostic manner.
    \item The approach reliably yields usable task instances (meaning $1+$ passing tests break).
    \item The approach is controllable; via each strategy's parameters, we can affect the quantity and quality of the generated bugs.
\end{enumerate}

\begin{example}[System prompt for generating bugs with an LM]
\small
You are a software developer doing chaos monkey testing.
Your job is to rewrite a function such that it introduces a logical bug that will break existing unit test(s) in a codebase.

To this end, some kinds of bugs you might introduce include: \\

\textcolor{red}{(Per inference call, only 3 of the following tips are randomly selected and shown)} \\
- Alter calculation order for incorrect results: Rearrange the sequence of operations in a calculation to subtly change the output (e.g., change (a + b) * c to a + (b * c)). \\
- Introduce subtle data transformation errors: Modify data processing logic, such as flipping a sign, truncating a value, or applying the wrong transformation function. \\
- Change variable assignments to alter computation state: Assign a wrong or outdated value to a variable that affects subsequent logic. \\
- Mishandle edge cases for specific inputs: Change handling logic to ignore or improperly handle boundary cases, like an empty array or a null input. \\
- Modify logic in conditionals or loops: Adjust conditions or loop boundaries (e.g., replace $<=$ with $<$) to change the control flow. \\
- Introduce off-by-one errors in indices or loop boundaries: Shift an index or iteration boundary by one, such as starting a loop at 1 instead of 0. \\
- Adjust default values or constants to affect behavior: Change a hardcoded value or default parameter that alters how the function behaves under normal use. \\
- Reorder operations while maintaining syntax: Rearrange steps in a process so the function produces incorrect intermediate results without breaking the code. \\
- Swallow exceptions or return defaults silently: Introduce logic that catches an error but doesn't log or handle it properly, leading to silent failures. \\

Tips about the bug-introducing task: \\
\textcolor{red}{(At inference time, tips are randomly shuffled)} \\
- It should not cause compilation errors. \\
- It should not be a syntax error. \\
- It should be subtle and challenging to detect. \\
- It should not modify the function signature. \\
- It should not modify the documentation significantly. \\
- For longer functions, if there is an opportunity to introduce multiple bugs, please do!"
- Please DO NOT INCLUDE COMMENTS IN THE CODE indicating the bug location or the bug itself. \\

Your answer should be formatted as follows: \\

  Explanation: $<$explanation$>$ \\

  Bugged Code: \\
  \texttt{```} \\
  $<$bugged\_code$>$ \\
  \texttt{```} \\
\end{example}
\captionof{figure}{
System prompt provided to an LM to generate bugs by modifying an existing, working function.
Text in \textcolor{red}{red} are not included at the actual prompt.
}
\label{fig:prompt_generate_bugs_with_lm}

\subsection{Generating with an LM}
\label{appx:bugs:generate:lm}
\begin{figure}[h]
    \centering
    \includegraphics[width=0.85\textwidth]{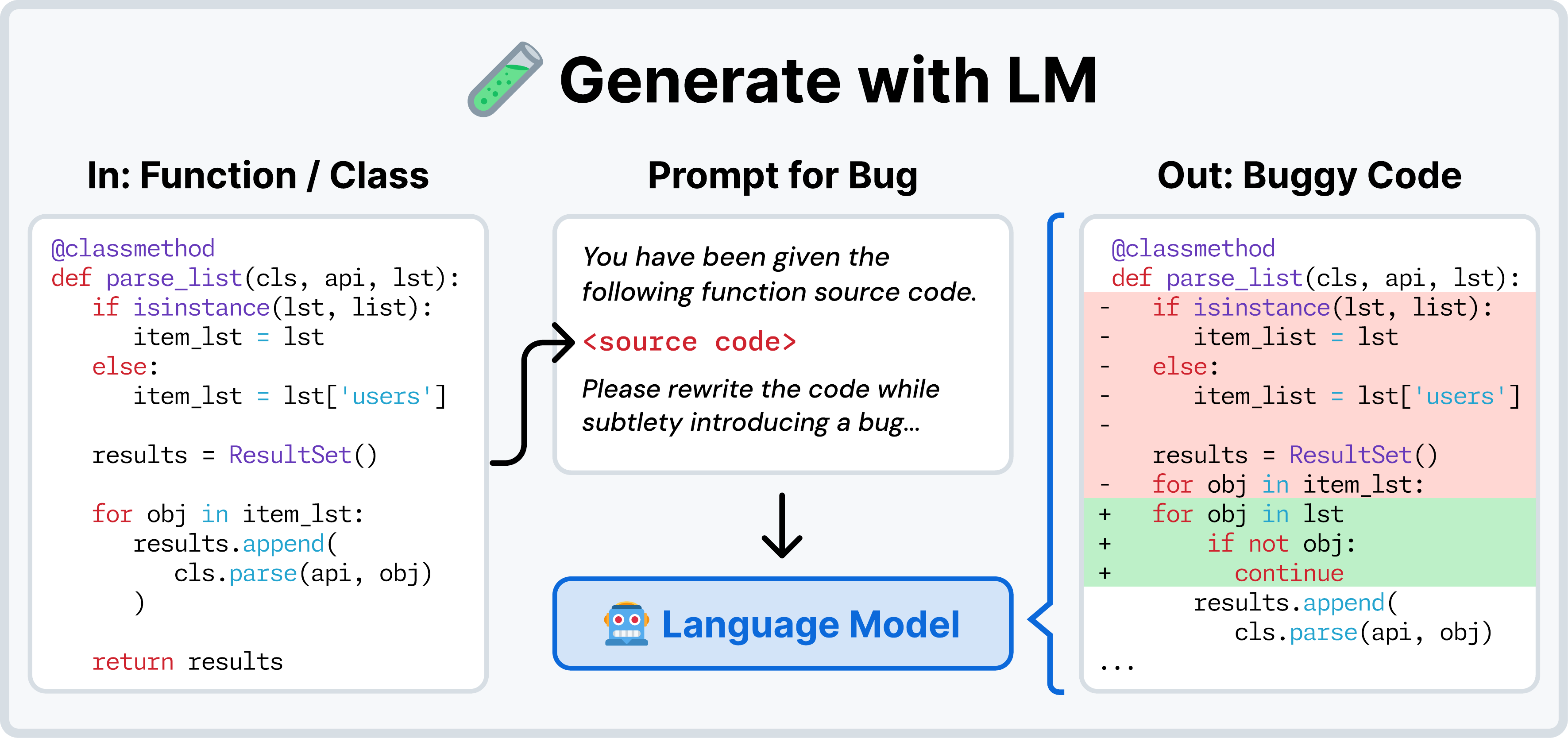}
    \caption{
    Workflow to generate bugs for a function or class with an LM.
    We first extract all functions or classes from a codebase, then enumerate across all candidates and prompt the LM to generate either a bug-laced rewrite or a re-implementation.
    }
    \label{fig:diagram-gen-with-lm}
\end{figure}

We describe our workflows for generating bugs with an LM.
For each function or class in a codebase, we prompt an LM to generate either a rewrite that introduces bugs or a complete re-implementation from scratch.
This strategy is illustrated in Figure~\ref{fig:diagram-gen-with-lm}.

\textbf{Modify existing functions.} Given a Python codebase, we use the \texttt{ast} library to identify all unique functions, excluding any functions found under a testing related directory (e.g. \texttt{tests}, \texttt{testing}).
Next, given a function, the LM is asked to write a new version that introduces logical, runtime bugs.
Within the prompt, shown in Figure~\ref{fig:prompt_generate_bugs_with_lm}, several suggestions of types of bugs along with a demonstration of a rewrite are provided.

\begin{example}[Prompts for reimplementing bugs with an LM]
\small
\textbf{System Prompt} \\
  You are a software developer and you have been asked to implement a function. \\

  You will be given the contents of an entire file, with one or more functions defined in it.
  Please implement the function(s) that are missing.
  Do NOT modify the function signature, including the function name, parameters, return types, or docstring if provided.
  Do NOT change any other code in the file.
  You should not use any external libraries. \\

\textbf{Task Instance Prompt} \\
  Please implement the function {func\_signature} in the following code: \\

  \{file\_src\_code\} \\

  Remember, you should not modify the function signature, including the function name, parameters, return types, or docstring if provided.
  Do NOT change any other code in the file.
  Format your output as: \\

  [explanation] \\

  \{func\_to\_write\}
\end{example}

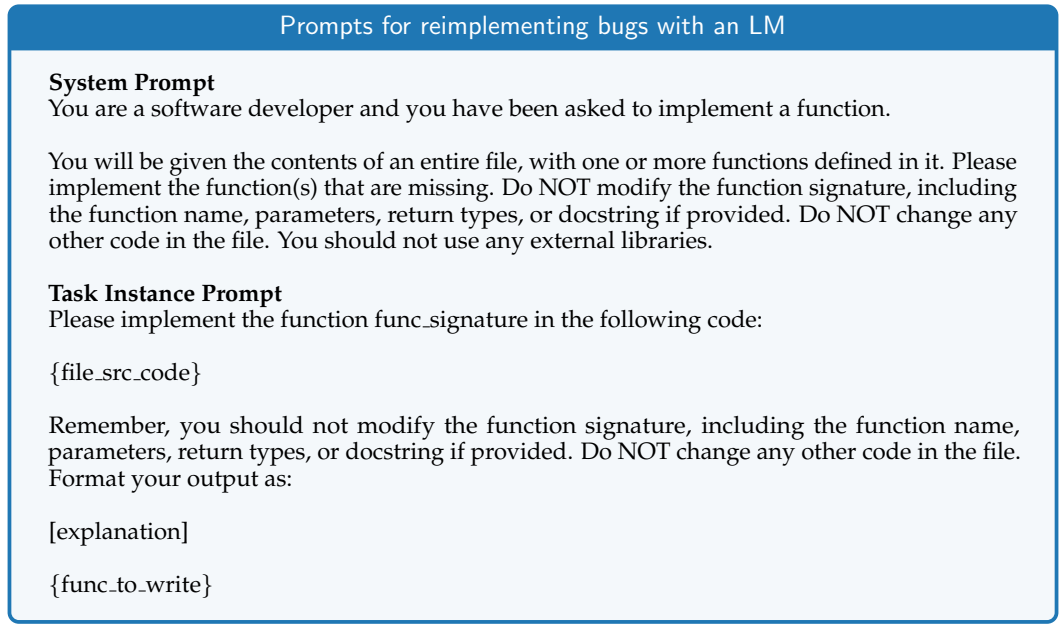
\captionof{figure}{
System prompt provided to an LM to generate bugs by re-implementing an existing target function.
\texttt{file\_src\_code} refers to the original source file minus the target function's original implementation.
\texttt{func\_to\_write} refers to the signature and docstring of the target function.
}
\label{fig:prompt_reimplement_func_with_lm}

In our experiments, we use OpenAI's o3 mini model~\citep{openai2024o3card} (\texttt{o3-mini-2025-01-31}) as the main base model for bug generation.
Based on our empirical observations of an LM's tendencies, we include several explicit guidelines in the prompt about what the rewrite should not do.
Notably, it is important to ask the LM to not generate any inline comments denoting the location of a bug; we observe that without explicitly specifying this, model generation outputs tend to have inline comments pointing out the bug.
We also want to avoid the complexities of identifying and removing such comments from a file diff representation.
Second, we state that rewrites causing compilation or syntax errors (e.g. undeclared variables, function definition modifications) should be avoided because such bugs are relatively trivial to solve.
We do not experiment extensively with different prompts or generating multiple buggy rewrites per function.

\textbf{Modify existing classes.}
This method involves a simple amendment to the function rewriting approach.
Instead of identifying unique functions (\texttt{ast.FunctionDef}), the codebase traversal logic instead looks for classes (\texttt{ast.ClassDef}).
Otherwise, all other aspects of the implementation are near identical to function rewriting, with minor changes to the prompt to make bug suggestions and the demonstration more class oriented.

\textbf{Rewrite existing functions.}
Instead of providing an LM with the original function, we explore an alternative strategy of asking an LM to re-implement a function from scratch.
Similar to above, we again use the \texttt{ast} library to identify all unique functions.
However, instead of directly asking for a bug, we remove the function's implementation, then prompt the LM with the entire file containing the function (minus the original implementation).
In the task description, we then explicitly ask for the LM to implement the function without changing the function signature.

\subsection{Procedural Modification}
\label{appx:bugs:generate:prod}
\begin{figure}[h]
    \centering
    \includegraphics[width=0.85\textwidth]{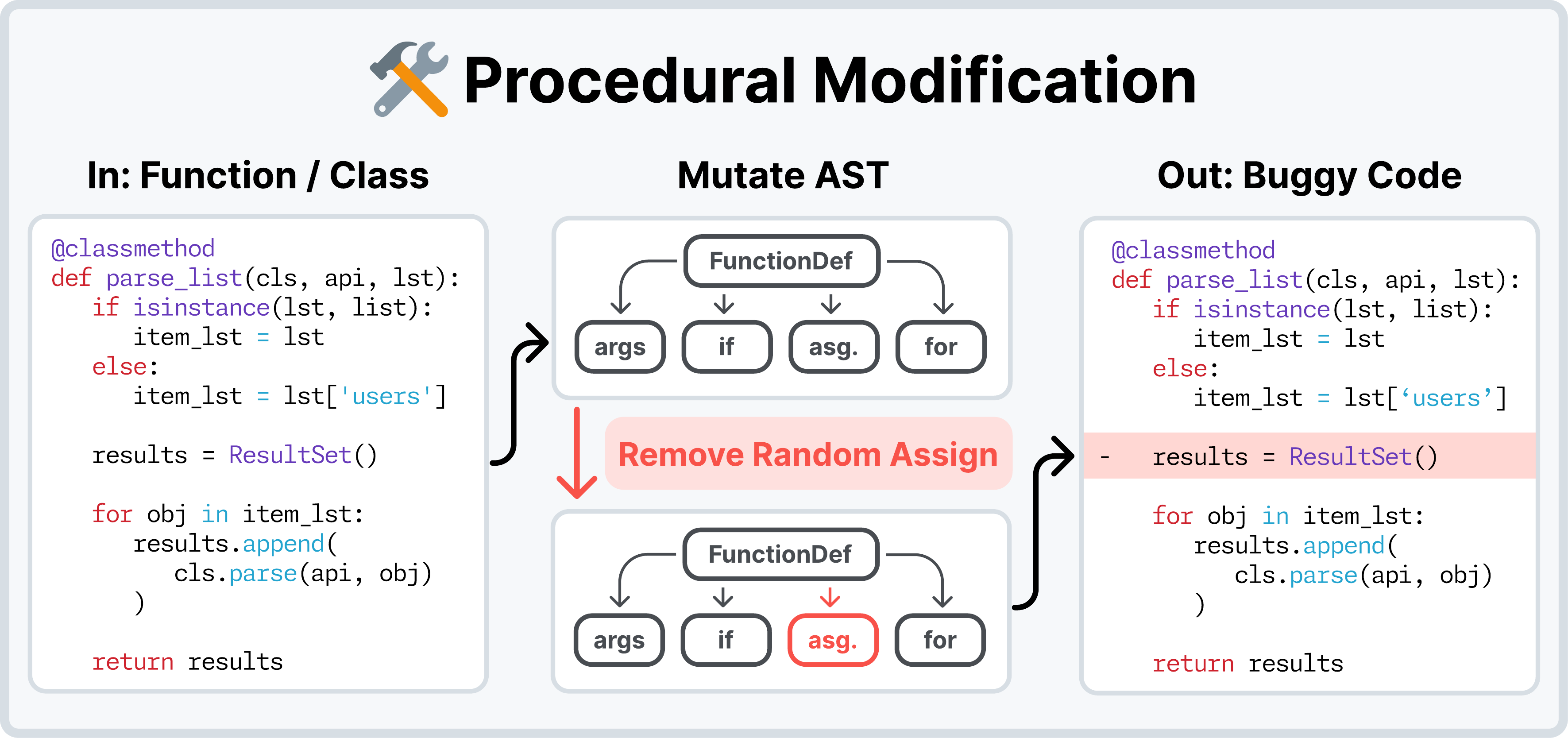}
    \caption{
    Workflow to generate bugs via procedural modifications.
    Per function/class, the source code is first convert into an \texttt{ast}.
    The modification then mutates the \texttt{ast} (e.g. removes an assignment statement).
    The \texttt{ast} is then converted back into source code with the specific modification introduced.
    }
    \label{fig:diagram-pm}
\end{figure}

We explore a zero-cost approach to create bugs by performing random modifications to the \texttt{ast} representation of a function or class.
A ``procedural modification" refers to a function that takes in an \texttt{ast} and applies a fixed transformation to it, such as removing a loop or swapping the blocks of an if/else clause.
This strategy is illustrated in Figure~\ref{fig:diagram-pm}.

Similar to the workflow for generating bugs with an LM, we first identify all functions or classes in a repository.
Per procedural modification, we first impose a set of criteria that filters out any candidates for which the modification would be impossible.
For instance, if the procedural modification removes a random conditional from a function, the modification's criteria will filter out any candidates that are not functions or do not have a conditional.
For the remaining candidates, the procedural modification is applied with controlled \texttt{likelihood}, where \texttt{likelihood} is a fraction indicating how often the procedural modification is applied within a candidate.
For example, if the procedural modification removes a random function with a \texttt{likelihood} of $0.5$, then for every conditional declared within the function, there is a $50$\% chance it gets removed.
We introduce \texttt{likelihood} so procedural modifications do not lead to changes that are too difficult.
Finally, the modified \texttt{ast} is converted back into source code.

Table~\ref{tab:pm_criteria} is a complete list of filtering criteria that is used for any procedural modification.
For the \texttt{filter\_min\_complexity} and \texttt{filter\_max\_complexity} criteria, we define a simple definition of ``complexity" as a sum of the number of conditional blocks, loops, boolean operators, exception handling blocks, and comparison operators in a function.
The purpose of \texttt{filter\_min\_complexity} is to remove both simple, uninteresting functions (e.g. getter, setter methods) from consideration.
\texttt{filter\_max\_complexity} is occasionally used to avoid changing long, monolithic functions.

\begin{table}[t]
    \centering
    \begin{tabular}{cl|l}
    \toprule
Index & Criteria & Description \\
    \midrule
1 & \texttt{filter\_functions} & Is the \texttt{ast} a function definition \\
2 & \texttt{filter\_classes} & Is the \texttt{ast} a class definition \\
3 & \texttt{filter\_classes\_has\_base} & Is the \texttt{ast} a class definition with parents \\
4 & \texttt{filter\_loops} & Does the \texttt{ast} contain a \texttt{For} or \texttt{While} loop? \\
5 & \texttt{filter\_conditionals} & Does the \texttt{ast} contain a conditional block? \\
6 & \texttt{filter\_assignments} & Is the \texttt{ast} a function def. with assignments? \\
7 & \texttt{filter\_wrappers} & Does the \texttt{ast} contain \texttt{try} or \texttt{with} blocks?  \\
8 & \texttt{filter\_if\_else} & Does the \texttt{ast} contain an \texttt{if-else} block? \\
9 & \texttt{filter\_operators} & Does the \texttt{ast} contain binary, boolean operators? \\
10 & \texttt{filter\_min\_complexity} & Is the \texttt{ast} $\geq$ a complexity score? \\
11 & \texttt{filter\_max\_complexity} & Is the \texttt{ast} $\leq$ a complexity score? \\
    \bottomrule
    \end{tabular}
    \caption{
    Pool of criteria used to filter for functions or classes with specific properties.
    Per procedural modification, a subset of these criteria is first used to filter functions and/or classes from a codebase.
    The modification is then run on the remainder.
    }
    \label{tab:pm_criteria}
\end{table}

Table~\ref{tab:pm_list} is an exhaustive list of all procedural modifications used to create bugs in a codebase.

\begin{table}[h]
    \centering
    \begin{tabular}{ll|ll}
\toprule
\multicolumn{2}{l|}{Procedural Modification} & Criteria & Description \\
\midrule
Class & Remove Functions & 2, 10 & Removes method(s) + reference(s). \\
      & Remove Parent & 3, 10 & Removes base class from class header. \\
      & Shuffle Methods & 2, 10 & Shuffles method definitions in a class. \\
\midrule
Control & Invert If/Else & 8 & Inverts the if-else bodies of a condition. \\
Flow    & Shuffle Lines & 11, 12 & Shuffles the lines of a function. \\
\midrule
Expressions & Change Constants & 1, 9, 10 & $\pm1$ to a constant numeric value.\\
            & Break Chains     & 1, 9, 10 & Removes operator(s), operator(s). \\
            & Swap Operands    & 1, 9, 10 & Mixes order of operands. \\
            & Change Operator  & 1, 9, 10 & Changes operator(s) (e.g. $+$ to $-$). \\
\midrule
Removal & Loops & 1, 4, 10        & Remove loops (e.g. \texttt{for}, \texttt{while}). \\
        & Conditionals & 1, 5, 10 & Remove conditionals (\texttt{if}). \\
        & Assignments & 1, 6, 10  & Remove assignment statements. \\
        & Wrappers & 1, 7, 10     & Remove exception (\texttt{try}), context (\texttt{with}). \\ 
\bottomrule
    \end{tabular}
    \caption{
    The $13$ procedural modification techniques we use to create bugs in a codebase.
    The ``Criteria" column contains indices referencing the corresponding filter defined in Table~\ref{tab:pm_criteria}.
    There are four informal categories --- Class, Control Flow, Expressions, Removal --- which indicates the general type of modification being made.
    }
    \label{tab:pm_list}
\end{table}

\subsection{Combine Bug Patches}
\label{appx:bugs:generate:combine}
We discuss the two strategies we use to combine bug patches from the same file or the same module.
In practice, we combine LM and procedurally generated bugs that have been validated successfully as usable task instances.

\begin{figure}[h]
    \centering
    \includegraphics[width=0.85\textwidth]{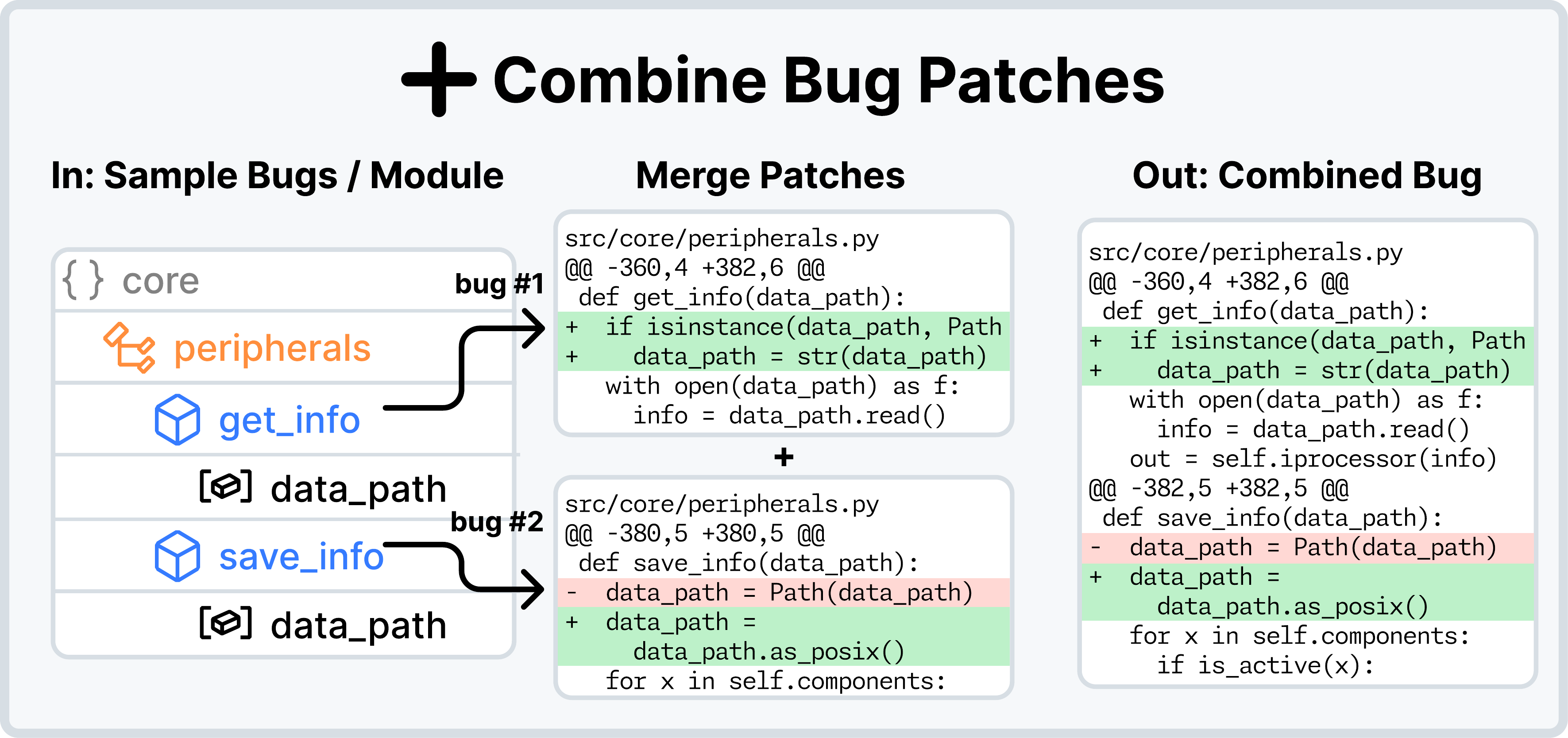}
    \caption{
    Workflow to generate bugs by combining bug patches.
    We take $n$ patches (generated using an LM or procedural modification), then sequentially apply each bug patch to the codebase.
    If all individual patches apply successfully, we save the resulting single patch which now represents all $n$ bugs combined.
    }
    \label{fig:diagram-combine}
\end{figure}

\textbf{From the same file.} If two or more functions are defined within a single file, this strategy merges the function-level bug patches together.
Given $n$ function-level bugs and $k$ as the number of bugs to combine, there are ${n\choose k}$ unique file-level candidate bug patches, which can be a large search space to cover.
To make the search space tractable, ensure no single function-level bug is repeatedly used, and generate instances that reliably have $1$+ Fail to Pass tests, we implement the following approach described in Algorithm~\ref{alg:combine_file}.

\begin{algorithm}
\begin{algorithmic}
\Require $codebase$, 
$bugs$; $num\_bugs$, $limit\_per\_file$; $max\_combos$
\Ensure $min\_bugs$ $\geq$ 2;
\State $max\_bugs$ $\geq$ $min\_bugs$;
\Procedure{CombineFileBugs}{}
\For{each $file$ in $codebase$}
    \State $file\_bugs$ $\gets$ bugs that apply to $file$
    \State $combinations$ $\gets$ get\_combos($file\_bugs$, $num\_bugs$, $max\_combos$)
    \For{each $combo$ in combinations}
        \State Apply $combo$ to $codebase$
        \If{success}
            \State Save $combo$ to disk
            \If{$limit\_per\_file$ reached}
                \State \textbf{break}
            \EndIf
            \State $combinations$ $\gets$ [c for c in $combinations$ if c $\cap$ $combo$ $=\emptyset$]
        \EndIf
    \EndFor
\EndFor
\EndProcedure
\end{algorithmic}
\caption{Combine multiple patches from the same file.}
\label{alg:combine_file}
\end{algorithm}

For each file in a codebase, we first identify the function-level bugs (or bug patches) that edit that file.
The pool of bugs we draw from have been \textit{validated}, meaning we have already ensured there is $1$+ Fail to Pass test(s) associated with the bug.
From these pool of \texttt{file\_bugs}, the \texttt{get\_combos} function then generates up to \texttt{max\_combos} sets of bugs, where the size of each set is \texttt{num\_bugs}.
For each \texttt{combo}, or set of bugs, the bugs are applied to the codebase one by one.
If all patches are successfully combined, this means they were successfully merged, and the merged patch, which consists of multiple function-level bugs, is saved and re-validated as a single bug.
Merging patches occasionally fails if there is an overlapping conflict between two files, akin to a merge conflict with \texttt{git}; this usually happens when a function is declared within another.
To ensure a function-level bug is only used once, any remaining bug sets in \texttt{combinations} using any patch in \texttt{combo} are removed.

The \texttt{limit\_per\_file} and \texttt{max\_combos} parameters prevent any one file from being over-represented and constrains an otherwise combinatorial large search space.
We run this algorithm across all codebase files, typically setting \texttt{num\_bugs}$=[2,4]$, \texttt{limit\_per\_file}$=3$, \texttt{max\_combos}$=40$.
Decreasing \texttt{num\_bugs} or increasing the other three parameters improves the yield.

\textbf{From the same module.} There are several ways one could imagine composing function-level bugs from multiple bugs, such as combining those that break the same test or have a programmatic relationship (e.g. function \texttt{a} calls function \texttt{b}).
We found a relatively straightforward and effective approach to be combining files that edit the same ``module".
By ``module" we are referring to a subdirectory within the source code (e.g. \texttt{sklearn/feature\_extraction}, \texttt{astropy/convolution}).
Out of all SWE-bench instances that edit $2$+ files, $75$\% modify files within the same submodule, suggesting a high degree of intra-module code changes.
The implementation for our approach is described in Algorithm~\ref{alg:combine_module}

\begin{algorithm}
\begin{algorithmic}
\Require $bugs$; $num\_bugs$; $limit\_per\_module$; $max\_combos$; $depth$
\Ensure $num\_bugs$ $\geq$ 2;
\Procedure{CombineModuleBugs}{}
\State $map\_path\_to\_bugs \gets \{\}$
    \For{each $bug$ in $bugs$}
        \State $path \gets$ get\_path\_from(bug)
        \State $map\_path\_to\_patches[path] \gets [bug]$
    \EndFor

    \State Collapse nested paths based on $depth$
    
    \ForAll{$(path, patches)$ in $map\_path\_to\_patches$}
        \State $combinations$ $\gets$ get\_combos(patches, $num\_bugs$, $max\_combos$)
        \For{each $combo$ in $combinations$}
            \State Apply $combo$ to $codebase$
            \If{success and num\_files\_changed(combo) $\geq 2$}
                \State Save $combo$ to disk
                \If{$limit\_per\_module$ reached}
                    \State \textbf{break}
                \EndIf
                \State $combinations$ $\gets$ [c for c in $combinations$ if c $\cap$ $combo$ $=\emptyset$]
            \EndIf
        \EndFor
    \EndFor
\EndProcedure
\caption{Combine multiple patches from the same module.}
\label{alg:combine_module}
\end{algorithmic}
\end{algorithm}

The implementation for this approach is similar to Algorithm~\ref{alg:combine_file} with two key changes.
First, we do not do file-by-file or folder-by-folder traversal.
Instead, using the diff patches, we create a dictionary \texttt{map\_path\_to\_bugs} that mimics the file structure of a codebase.
For example, if \texttt{bug} modifies path \texttt{a/b/c/d.py}, it is represented as \texttt{map\_path\_to\_bugs[a][b][c][d.py]} $=$ \texttt{[bug]}.
Additional bugs that modify the same path are appended to the list.
Since every bug is a function-level bug, there will never be a bug registered in multiple lists.
We then ``collapse" up to \texttt{depth} indices.
So for instance, at \texttt{depth} $=3$, the above data structure is collapsed into \texttt{map\_path\_to\_bugs[a/b/c][d.py]} $=$ \texttt{[bug]}.
Finally, any nested dictionaries are collapsed into a single list of patches (e.g. \texttt{map\_path\_to\_bugs[a/b/c]} $=$ \texttt{[bug]}).
Mirroring the procedure in Algorithm~\ref{alg:combine_file}, we then iterate across this dictionary's values (lists of bugs).
Second, we only save patches that modify $2+$ files; aggregate bugs (represented by \texttt{combo}) modifying a single file are not considered.

Again, we run this strategy across all $100$ repositories, with parameters \texttt{num\_bugs}$=[2,5]$, \texttt{limit\_per\_module}$=10$, \texttt{max\_combos}$=100$, and \texttt{depth}$=2$.
Reducing \texttt{num\_bugs}, \texttt{depth} and increasing the other parameters yields more bugs.
We choose a \texttt{depth} of $2$ because empirically, we find that meaningful modules are usually declared as immediate sub-folders of the main source code folder (e.g. in \texttt{sklearn/feature\_extraction}, \texttt{sklearn} is the source code folder while \texttt{feature\_extraction} is the module).
A shallower depth leads to less meaningful groupings, while yield decreases significantly for every increased level of depth, particularly for smaller repositories.

\subsection{Pull Request Mirroring}
\label{appx:bugs:generate:pr}
\begin{figure}[h]
    \centering
    \includegraphics[width=0.85\textwidth]{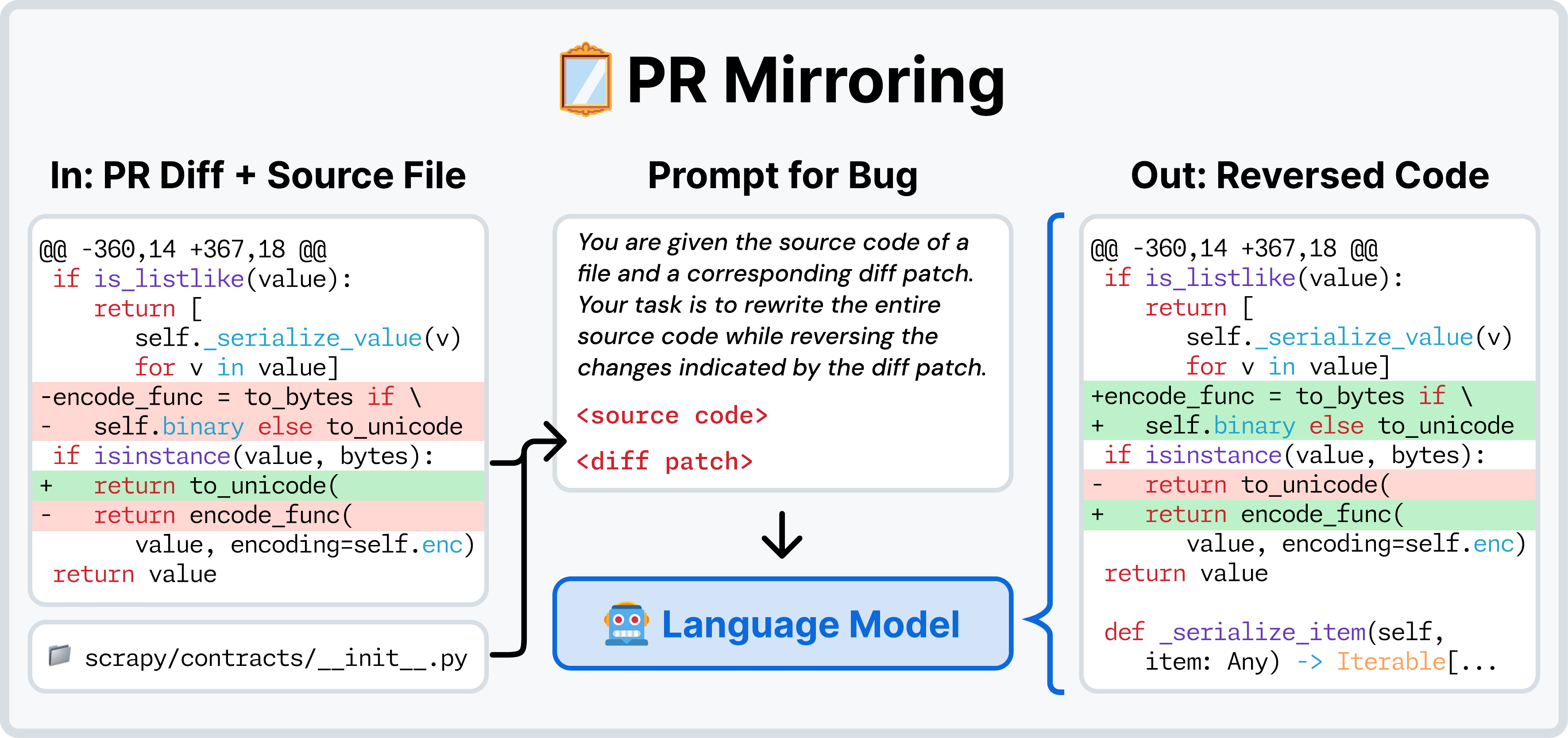}
    \caption{
    Workflow to generate bugs by reverting changes made in the diff patch corresponding to a real GitHub pull request (PR).
    Given the patch and the files modified by the patch, we prompt the LM to generate a complete rewrite of each file that \textit{reverses} the changes made in the PR.
    The changes are applied to the codebase, and we extract the patch, which now captures the reversal of the PR changes.
    }
    \label{fig:diagram-pr-mirror}
\end{figure}

We finally discuss the fourth and last strategy for generating bugs - mirroring real world pull requests (PR).
We visualize this process in Figure~\ref{fig:diagram-pr-mirror}.

\textbf{Why use an LM?}
When we initially implemented this approach, we attempted to directly perform a \texttt{git apply --reverse [patch]} on the codebase.
However, for the large majority of patches, this fails.
We performed troubleshooting by inspecting $100$ PR patches on the \texttt{sqlfluff/sqlfluff} repository, leading us to two observations.
\begin{enumerate}
    \item The majority of these PRs reflect changes that remain present in the codebase today (making the bug creation promising).
    \item However, many patches can not be reversed because the exact location (e.g. lines, file) of the relevant code changed because of other changes.
\end{enumerate}
Therefore, we employ LMs to perform patch reversal, and find that reasoning models (e.g. \texttt{o3-mini}~\citep{openai2024o3card}) are particularly effective.

\textbf{Description of method.} We follow SWE-bench's methodology for crawling PRs created January 1st, 2023 and onwards, with minor and arbitrary exceptions for some repositories where we crawl older PRs as well.
Per PR, we iterate across the file(s) changed by the patch.
Per file, we prompt an LM with the file-specific changes from the patch along with the file's source code in the current state of the repository (\textit{not} the repository's state corresponding to when the PR was applied, referred to as the \texttt{base\_commit} in SWE-bench).
The LM is asked to generate a rewrite of the file that reverts the changes reflected in the PR.
We aggregate the changes across all file(s) into a single patch.

Because we are interested in problems that our expert trajectory generation method (SWE-agent + Claude 3.7 Sonnet) has a chance of solving, we do not attempt to reproduce PRs that change more than $8$ files.
This constraint is imposed because no SWE-bench instance that edits more than $6$ files has ever been solved~\citep{jimenez2024leaderboard}.

\textbf{How well does PR mirroring work?} We scrape the PRs corresponding to $100$ randomly selected SWE-bench task instances from the \texttt{django/django} GitHub repository and attempt to recreate these task instances with \bugs{}'s collection process.
We successfully recovered $92$ of $100$ task instances.
Of these, $84$ break identical F2P test(s), with the remaining $8$ breaking a subset because some tests were removed over time.
This sanity check gives us confidence that the PR mirroring strategy lives up to its name.

\textbf{Comparison to SWE-bench.} This approach has several benefits and drawbacks compared to SWE-bench's collection pipeline.
First, it removes the need to create instance-specific Docker images --- all PRs are mirrored against the same version of a repository.
This also implies that there is no need to write installation specifications for past versions of a repository, which is typically the most laborious step in task construction with SWE-bench.
Finally, this strategy also allows us to loosen the requirements on what PRs we attempt to convert into a task instance.
In SWE-bench, the core requirements for what PRs to attempt to convert into a task instance include:
\begin{enumerate}
    \setlength\itemsep{0em}
    \item It must edit $1+$ code files (e.g. not just \texttt{.md}, \texttt{.rst} files).
    \item It must reference $1+$ GitHub issues, which serves as the problem statement.
    \item It must edit $1+$ testing related files ($1+$ files with a \texttt{test}-adjacent keyword in it).
\end{enumerate}

With this collection strategy and \bugs{}'s focus on training data, the second and third requirements are no longer necessary.
If there is no associated issue, issue text can simply be generated.
If the patch does not contain any testing related changes, this is tolerable, as the validation stage will determine whether the PR breaks any tests.
With these considerations, we purport that \bugs{}'s PR mirroring strategy can re-purpose a higher percentage of real world code changes for training purposes.

The main downside is that the rest of the repository is out of sync with the state of the codebase when the PR was applied.
As a result, it's possible that changes in the behavior of the rest of the codebase may affect the issue's reproducibility or the accuracy of the issue description (e.g. line numbers referenced in the issue text are likely somewhat off with respect to the codebase).
However, a simple mitigation for this is to create a Docker image for a repository at an earlier commit that's closer to the original creation date of the issue.
While we do not carry out a targeted experiment, we hypothesize that using \bugs{}, we would be able to reproduce SWE-bench entirely with $10$x less human hours with an estimated $2294$ x \$$0.055$ = \$$126.17$ in costs.
\section{Dataset Statistics}
\label{appx:dataset}
We present additional breakdowns and analyses of the \bugs{} dataset, focusing on the kinds of repositories and bugs that are represented.

\textbf{Repository categorization.}
We present an exhaustive list of repositories used in \bugs{} in Table~\ref{tab:summary_of_repos}.
We categorize the repositories into seven general buckets: Data Parsing and Transformation ($39$), Web \& API Development ($11$), Code Quality \& Testing ($12$), Visualization \& Presentation ($8$), System Tools \& Protocols ($17$), Natural Language Processing ($7$), and Miscellaneous ($6$).
The categorizations were performed by first, determining an appropriate set of categories based on manual inspection supported by the descriptions and GitHub topics associated with each repository.
After settling upon the buckets, we asked GPT-4o to provide a label based on the repository's metadata and \texttt{README} dump.
\bugs{} represents a wider and more variegated coverage of software tools and applications compared to any prior works.

\begin{CJK*}{UTF8}{gbsn}
\begin{longtable}{p{0.3\linewidth} | p{0.65\linewidth}}
\toprule
\textbf{Repository} & \textbf{Description} \\
\midrule
\endfirsthead

\toprule
\textbf{Repository} & \textbf{Description} \\
\midrule
\endhead

\midrule
\multicolumn{2}{r}{\textit{Continued on next page}} \\
\endfoot

\bottomrule
\endlastfoot

\multicolumn{2}{c}{\textit{Code Quality and Testing}} \\
\midrule
PyCQA/flake8 & flake8 is a python tool that glues together pycodestyle, pyflakes, mccabe, and third-party plugins to check the style and quality of some python code. \\
Suor/funcy & A fancy and practical functional tools \\
adrienverge/yamllint & A linter for YAML files. \\
agronholm/typeguard & Run-time type checker for Python \\
cknd/stackprinter & Debugging-friendly exceptions for Python \\
cool-RR/PySnooper & Never use print for debugging again \\
getmoto/moto & A library that allows you to easily mock out tests based on AWS infrastructure. \\
pylint-dev/astroid & A common base representation of python source code for pylint and other projects \\
pytest-dev/iniconfig None
pytest-dev/iniconfig & None \\
python/mypy & Optional static typing for Python \\
pyupio/safety & Safety checks Python dependencies for known security vulnerabilities and suggests the proper remediations for vulnerabilities detected. \\
pyutils/line\_profiler & Line-by-line profiling for Python \\
rubik/radon & Various code metrics for Python code \\
spulec/freezegun & Let your Python tests travel through time \\
sqlfluff/sqlfluff & A modular SQL linter and auto-formatter with support for multiple dialects and templated code. \\
\midrule
\multicolumn{2}{c}{\textit{Data Parsing and Transformation}} \\
\midrule
alecthomas/voluptuous & CONTRIBUTIONS ONLY: Voluptuous, despite the name, is a Python data validation library. \\
andialbrecht/sqlparse & A non-validating SQL parser module for Python \\
buriy/python-readability & fast python port of arc90's readability tool, updated to match latest readability.js! \\
burnash/gspread & Google Sheets Python API \\
chardet/chardet & Python character encoding detector \\
cloudpipe/cloudpickle & Extended pickling support for Python objects \\
dask/dask & Parallel computing with task scheduling \\
datamade/usaddress & :us: a python library for parsing unstructured United States address strings into address components \\
davidhalter/parso & A Python Parser \\
erikrose/parsimonious & The fastest pure-Python PEG parser I can muster \\
facelessuser/soupsieve & A modern CSS selector implementation for BeautifulSoup \\
gawel/pyquery & A jquery-like library for python \\
google/textfsm & Python module for parsing semi-structured text into python tables. \\
gruns/furl & �� URL parsing and manipulation made easy. \\
gweis/isodate & ISO 8601 date/time parser \\
hukkin/tomli & A lil' TOML parser \\
jawah/charset\_normalizer & Truly universal encoding detector in pure Python \\
john-kurkowski/tldextract & Accurately separates a URL’s subdomain, domain, and public suffix, using the Public Suffix List (PSL). \\
joke2k/faker & Faker is a Python package that generates fake data for you. \\
jsvine/pdfplumber & Plumb a PDF for detailed information about each char, rectangle, line, et cetera — and easily extract text and tables. \\
kayak/pypika & PyPika is a python SQL query builder that exposes the full richness of the SQL language using a syntax that reflects the resulting query. PyPika excels at all sorts of SQL queries but is especially useful for data analysis. \\
keleshev/schema & Schema validation just got Pythonic \\
kennethreitz/records & SQL for Humans™ \\
kurtmckee/feedparser & Parse feeds in Python \\
lepture/mistune & A fast yet powerful Python Markdown parser with renderers and plugins. \\
madzak/python-json-logger & Json Formatter for the standard python logger \\
mahmoud/glom & ☄️ Python's nested data operator (and CLI), for all your declarative restructuring needs. Got data? Glom it! ☄️ \\
marshmallow-code/marshmallow & A lightweight library for converting complex objects to and from simple Python datatypes. \\
martinblech/xmltodict & Python module that makes working with XML feel like you are working with JSON \\
matthewwithanm/python-markdownify & Convert HTML to Markdown \\
mewwts/addict & The Python Dict that's better than heroin. \\
mido/mido & MIDI Objects for Python \\
modin-project/modin & Modin: Scale your Pandas workflows by changing a single line of code \\
mozilla/bleach & Bleach is an allowed-list-based HTML sanitizing library that escapes or strips markup and attributes \\
msiemens/tinydb & TinyDB is a lightweight document oriented database optimized for your happiness :) \\
pandas-dev/pandas & Flexible and powerful data analysis / manipulation library for Python, providing labeled data structures similar to R data.frame objects, statistical functions, and much more \\
pdfminer/pdfminer.six & Community maintained fork of pdfminer - we fathom PDF \\
pudo/dataset & Easy-to-use data handling for SQL data stores with support for implicit table creation, bulk loading, and transactions. \\
pydantic/pydantic & Data validation using Python type hints \\
pydata/patsy & Describing statistical models in Python using symbolic formulas \\
pydicom/pydicom & Read, modify and write DICOM files with python code \\
pygments/pygments & Pygments is a generic syntax highlighter written in Python \\
pyparsing/pyparsing & Python library for creating PEG parsers \\
python-jsonschema/jsonschema & An implementation of the JSON Schema specification for Python \\
python-openxml/python-docx & Create and modify Word documents with Python \\
r1chardj0n3s/parse & Parse strings using a specification based on the Python format() syntax. \\
scanny/python-pptx & Create Open XML PowerPoint documents in Python \\
scrapy/scrapy & Scrapy, a fast high-level web crawling \& scraping framework for Python. \\
seperman/deepdiff & DeepDiff: Deep Difference and search of any Python object/data. DeepHash: Hash of any object based on its contents. Delta: Use deltas to reconstruct objects by adding deltas together. \\
sloria/environs & simplified environment variable parsing \\
sunpy/sunpy & SunPy - Python for Solar Physics \\
tkrajina/gpxpy & gpx-py is a python GPX parser. GPX (GPS eXchange Format) is an XML based file format for GPS tracks. \\
tobymao/sqlglot & Python SQL Parser and Transpiler \\
un33k/python-slugify & Returns unicode slugs \\
\midrule
\multicolumn{2}{c}{\textit{Machine Learning and AI}} \\
\midrule
facebookresearch/fvcore & Collection of common code that's shared among different research projects in FAIR computer vision team. \\
facebookresearch/hydra & Hydra is a framework for elegantly configuring complex applications \\
HIPS/autograd & Efficiently computes derivatives of NumPy code. \\
iterative/dvc & �� Data Versioning and ML Experiments \\
jaraco/inflect & Correctly generate plurals, ordinals, indefinite articles; convert numbers to words \\
life4/textdistance & �� Compute distance between sequences. 30+ algorithms, pure python implementation, common interface, optional external libs usage. \\
luozhouyang/python-string-similarity & A library implementing different string similarity and distance measures using Python. \\
Mimino666/langdetect & Port of Google's language-detection library to Python. \\
mozillazg/python-pinyin & 汉字转拼音(pypinyin) \\
pndurette/gTTS & Python library and CLI tool to interface with Google Translate's text-to-speech API \\
Project-MONAI/MONAI & AI Toolkit for Healthcare Imaging \\
seatgeek/thefuzz & Fuzzy String Matching in Python \\
vi3k6i5/flashtext & Extract Keywords from sentence or Replace keywords in sentences. \\
\midrule
\multicolumn{2}{c}{\textit{System Tools and Protocols}} \\
\midrule
agronholm/exceptiongroup & Backport of PEP 654 (exception groups) \\
aio-libs/async-timeout & asyncio-compatible timeout class \\
arrow-py/arrow & �� Better dates \& times for Python \\
borntyping/python-colorlog & A colored formatter for the python logging module \\
cantools/cantools & CAN bus tools. \\
conan-io/conan & Conan - The open-source C and C++ package manager \\
cookiecutter/cookiecutter & A cross-platform command-line utility that creates projects from cookiecutters (project templates), e.g. Python package projects, C projects. \\
dbader/schedule & Python job scheduling for humans. \\
gruns/icecream & �� Never use print() to debug again. \\
jd/tenacity & Retrying library for Python \\
mahmoud/boltons & �� Like builtins, but boltons. 250+ constructs, recipes, and snippets which extend (and rely on nothing but) the Python standard library.  Nothing like Michael Bolton. \\
oauthlib/oauthlib & A generic, spec-compliant, thorough implementation of the OAuth request-signing logic \\
pallets/click & Python composable command line interface toolkit \\
paramiko/paramiko & The leading native Python SSHv2 protocol library. \\
pexpect/ptyprocess & Run a subprocess in a pseudo terminal \\
pyasn1/pyasn1 & Generic ASN.1 library for Python \\
pyca/pyopenssl & A Python wrapper around the OpenSSL library \\
python-hyper/h11 & A pure-Python, bring-your-own-I/O implementation of HTTP/1.1 \\
python-trio/trio & Trio – a friendly Python library for async concurrency and I/O \\
rustedpy/result & NOT MAINTAINED - A simple Rust like Result type for Python 3. Fully type annotated. \\
termcolor/termcolor & ANSI color formatting for output in terminal \\
theskumar/python-dotenv & Reads key-value pairs from a .env file and can set them as environment variables. It helps in developing applications following the 12-factor principles. \\
tox-dev/pipdeptree & A command line utility to display dependency tree of the installed Python packages \\
\midrule
\multicolumn{2}{c}{\textit{Visualization and Presentation}} \\
\midrule
amueller/word\_cloud & A little word cloud generator in Python \\
lincolnloop/python-qrcode & Python QR Code image generator \\
prettytable/prettytable & Display tabular data in a visually appealing ASCII table format \\
pwaller/pyfiglet & An implementation of figlet written in Python \\
rsalmei/alive-progress & A new kind of Progress Bar, with real-time throughput, ETA, and very cool animations! \\
weaveworks/grafanalib & Python library for building Grafana dashboards \\
\midrule
\multicolumn{2}{c}{\textit{Web and API Development}} \\
\midrule
Cog-Creators/Red-DiscordBot & A multi-function Discord bot \\
Knio/dominate & Dominate is a Python library for creating and manipulating HTML documents using an elegant DOM API.  It allows you to write HTML pages in pure Python very concisely, which eliminate the need to learn another template language, and to take advantage of the more powerful features of Python. \\
alanjds/drf-nested-routers & Nested Routers for Django Rest Framework \\
benoitc/gunicorn & gunicorn 'Green Unicorn' is a WSGI HTTP Server for UNIX, fast clients and sleepy applications. \\
bottlepy/bottle & bottle.py is a fast and simple micro-framework for python web-applications. \\
django-money/django-money & Money fields for Django forms and models. \\
django/channels & Developer-friendly asynchrony for Django \\
django/daphne & Django Channels HTTP/WebSocket server \\
encode/starlette & The little ASGI framework that shines. �� \\
getnikola/nikola & A static website and blog generator \\
graphql-python/graphene & GraphQL framework for Python \\
marshmallow-code/apispec & A pluggable API specification generator. Currently supports the OpenAPI Specification (f.k.a. the Swagger specification).. \\
marshmallow-code/webargs & A friendly library for parsing HTTP request arguments, with built-in support for popular web frameworks, including Flask, Django, Bottle, Tornado, Pyramid, webapp2, Falcon, and aiohttp. \\
pallets/jinja & A very fast and expressive template engine. \\
pallets/markupsafe & Safely add untrusted strings to HTML/XML markup. \\
tornadoweb/tornado & Tornado is a Python web framework and asynchronous networking library, originally developed at FriendFeed. \\
tweepy/tweepy & Twitter for Python! \\
\end{longtable}
\end{CJK*}
\label{tab:summary_of_repos}

\subsection{Bug Generation Statistics}
\label{appx:dataset:bug_gen_stats}
We provide extensive details about different aspects of each of the bug generation strategies, including the yield rates, labor/monetary costs, and dataset characterizations.

\textbf{Yield rates.}
In Table~\ref{tab:per_bug_type_yield_rate}, we provide the yield rates for each bug generation method across all repositories in \bugs{}.
In general, we find that the PR Mirroring has the lowest yield rate at $13.18$\% (although this rate is somewhat higher than SWE-bench's yield rate of $2294/93139 = 2.46$\%).
For using LMs to generate bugs, modifying functions to introduce bugs intentionally has a higher yield than asking LMs to perform a best-effort rewrite.
The efficacy of Procedural Modifications varies by strategy.
For instance, shuffling the functions declared in a class only breaks existing test(s) $1.93$\% of the time, but inverting a conditional will lead to a task instance for $47.04$\% of modifications.
Finally, combining bug patches has an extremely high yield rate - this is to be expected because we only attempt to combine bug patches that have been validated as usable task instances breaking $1+$ tests.

\begin{table}[h]
    \centering
    \begin{tabular}{l|rrrr}
\toprule
Strategy & \# Repos & \# Candidates & \# Instances & Yield Rate \\
\midrule
Combine (file) & 124 & 6020 & 5865 & 97.43\% \\
Combine (module) & 65 & 4396 & 4227 & 96.16\% \\
LM (Modify) & 108 & 31950 & 17887 & 55.98\% \\
LM (Rewrite) & 128 & 11908 & 4173 & 35.04\% \\
PR Mirroring & 108 & 6934 & 2344 & 33.8\% \\
Procedural (Class Rm Base) & 103 & 1401 & 463 & 33.05\% \\
Procedural (Class Rm Funcs) & 103 & 2506 & 1180 & 47.09\% \\
Procedural (Class Shuffle Funcs) & 103 & 2504 & 47 & 1.88\% \\
Procedural (Ctrl Invert If) & 105 & 4695 & 2321 & 49.44\% \\
Procedural (Ctrl Shuffle) & 104 & 9055 & 4015 & 44.34\% \\
Procedural (Op Break Chains) & 71 & 747 & 225 & 30.12\% \\
Procedural (Op Change Const) & 77 & 723 & 257 & 35.55\% \\
Procedural (Op Change) & 81 & 1507 & 450 & 29.86\% \\
Procedural (Op Swap) & 87 & 2141 & 483 & 22.56\% \\
Procedural (Remove Assign) & 121 & 5470 & 2661 & 48.65\% \\
Procedural (Remove Cond) & 120 & 5288 & 2311 & 43.7\% \\
Procedural (Remove Loop) & 110 & 1945 & 860 & 44.22\% \\
Procedural (Remove Wrapper) & 80 & 884 & 368 & 41.63\% \\
\midrule
All & 129 & 100074 & 50137 & 50.1\% \\
\bottomrule
    \end{tabular}
    \caption{
    Yield rates for different bug generation strategies covered in Section~\ref{appx:bugs:generate}.
    We show the number of repositories that each strategy was run on, the number of bug candidates generated by each strategy, and the number of instances, or the number of candidates that were validated to have $1+$ Fail to Pass test.
    The yield rate for 
    }
    \label{tab:per_bug_type_yield_rate}
\end{table}

The number of repositories captured by each bug generation technique varies due to each strategy's specific preconditions, which at times may not be effective for some repositories.
For instance, the \textit{Procedural (Class *)} set of methods only mutates Python classes.
This strategy is fruitless for the minority of \bugs{} repositories that do not define any classes.
The \textit{Procedural (Op Break Chains)} method randomly removes operations and operands from expressions with two or more operations (e.g. $a+b+c \rightarrow a+b$) --- such expressions are not always present in \bugs{} repositories.

The collective yield rate across \bugs{}'s bug generation strategies is significantly higher than SWE-bench's collection strategy.

\begin{table}[ht]
    \centering
    \begin{tabular}{l|c}
\toprule
Yield Rate & \# of Repositories \\
\midrule
$0$-$25$\%   & $10$  \\
$25$-$50$\%  & $31$ \\
$50$-$75$\%  & $60$ \\
$75$-$100$\% & $27$ \\
\bottomrule
    \end{tabular}
    \caption{
    Yield rates for different repositories represented in \bugs{}.
    }
    \label{tab:per_repo_yield_rate}
\end{table}

The yield rate also varies with respect to the repository it is being applied to.
We provide a summary of yield rates by repository in Table~\ref{tab:per_repo_yield_rate}.
We generally observe that lower test coverage correlates with a lower yield rate.

\textbf{Dataset characterizations.}
In Table~\ref{tab:per_bug_type_stats}, we provide statistics about the validated task instances produced by different bug generation strategies.
Our work's LM-based strategies rewrite one function in one file.
Procedural modifications will also only change one file, but depending on the strategy, $1+$ functions or classes may be changed.
Combining multiple patches from the same file always produces a patch with $2+$ functions edited.
Combining across modules produces a patch with $2+$ files edited.
The targeted nature of each of the bug creation strategies is reflected in the typical number of functions and files that the bugs produced by each strategy edits.

\begin{table}[h]
    \centering
    \begin{tabular}{l|rrrrr}
\toprule
Strategy & \# Instances & \# F2P & $\Delta$ Lines & $\Delta$ Functions & $\Delta$ Files \\
\midrule
Combine      & 10092 & 15 (5-48) & 19 (12-36) & 2 (2-3) & 1 (1-2) \\
LM           & 22060 & 4 (1-17) & 6 (3-15) & 1 (1-1) & 1 (1-1) \\
PR Mirroring & 2344 & 3 (1-14) & 20 (8-55) & 2 (2-4) & 1 (1-2) \\
Procedural   & 15641 & 7 (2-32) & 7 (5-15) & 1 (1-1) & 1 (1-1) \\
\bottomrule
    \end{tabular}
    \caption{Statistics for attributes of a \bugs{} task instance across different bug generation strategies, reported as \textit{median (IQR)}, where IQR is the inter-quartile range (25th–75th percentile).
    }
    \label{tab:per_bug_type_stats}
\end{table}

\begin{figure}[h]
    \centering
    \includegraphics[width=\textwidth]{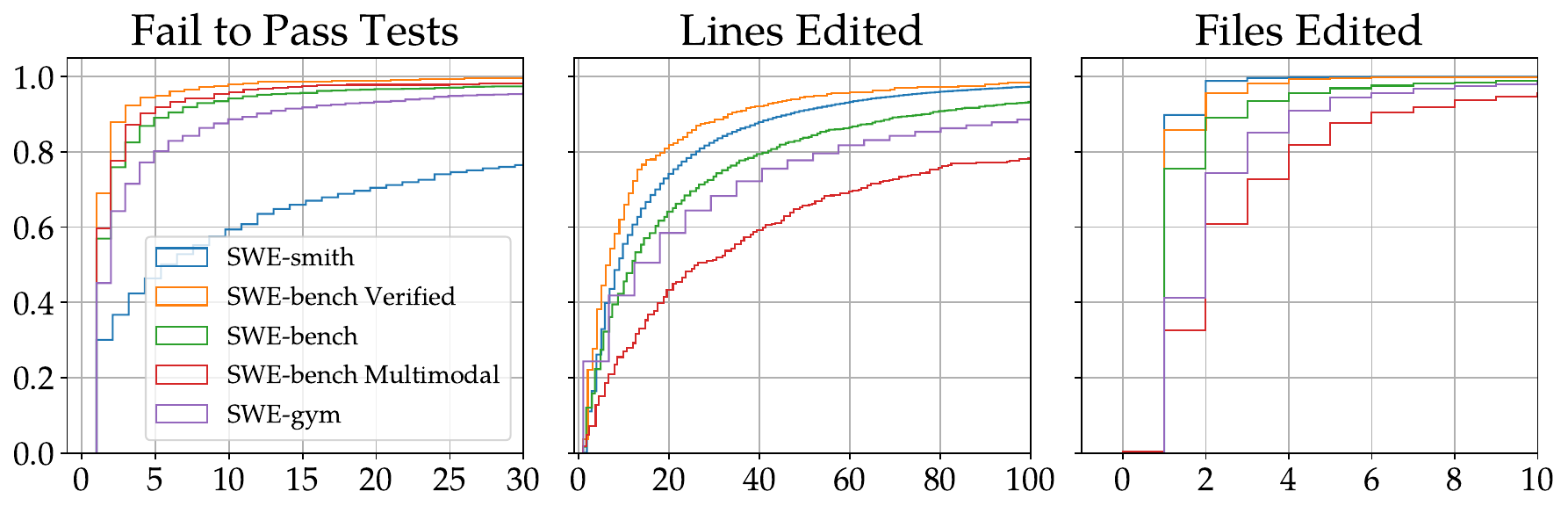}
    \caption{
    Comparison of cumulative distributions for Fail-to-Pass tests along with the lines and files edited by the gold patch across \bugs{} and four SWE-bench style datasets.
    }
    \label{fig:dataset_comparison}
\end{figure}

In Figure~\ref{fig:dataset_comparison}, we show the distributions for different attributes of \bugs{} compared to other SWE-bench style datasets.
Compared to prior works, there is a much higher proportion of task instances with more than one Fail-to-Pass test.
For any one repository, we find that \bugs{} task instances collectively cause failures for a much higher percentage of the testing suit than other datasets; a potential benefit of this is that training on \bugs{} based trajectories may expose models to a much broader set of functionalities in a codebase.
The number of lines and files edited by \bugs{} task instances is highly similar to the trend lines for SWE-bench Verified.

We note that unlike other datasets, the trend line of \bugs{} task instances is ``adjustable".
In other words, the Figure~\ref{fig:dataset_comparison} distributions are a capture of the task instances provided in this release of \bugs{}.
However, because of \bugs{}'s flexible bug creation techniques, the distribution can be ``shaped" if needed.
For instance, generating more task instances using the bug patch combination method would shift all three curves in Figure~\ref{fig:dataset_comparison}.
We make this point to highlight the fact that the attributes of SWE-bench task instances are, in a sense, constrained by real world software development behavior.
On the other hand, \bugs{} can be used to break tests and code that may not be reflected at all in any existing pull request.
In this sense, we argue that LMs trained on \bugs{} have better ``exposure" to a codebase compared to exclusively training on pull requests.

\textbf{Continuation of scaling execution environments.}
The validation and evaluation procedures for \bugs{} deviate slightly from SWE-bench's harnesses.
The main reasons for these differences can largely be attributed to the granularity of installation specifications.
In SWE-bench, each task instance corresponds to a unique base commit, with additional \texttt{version} and \texttt{environment\_setup\_commit} keys needed as indirection for mapping an instance to the correct set of installation and testing instructions.
Across time, the continuous evolution of a repository and its dependencies make for an incredibly high degree of variability in how a repository should be installed correctly.
To solve this variability, the community has resorted to creating an image per task instance, as done in~\citet{chowdhuryintroducing}.
Therefore, for $2294$ SWE-bench task instances, there are $2294$ unique Docker images, each at a size of at least several gigabytes ($\sim5$-$6$ GBs).

On the other hand, the simplicity and scalability of \bugs{}'s design allows one to support many task instances with comparatively much fewer Docker images.
As mentioned above, installation and testing procedures are (repository, commit) specific.
Therefore, when bugs are generated from each (repository, commit), all bugs can be reproduced and tested successfully from the same Docker image.
In other words, if I generate $100$ bugs for a repository at some commit, instead of $100$ Docker images, only a single Docker image is required to run inference on any of the $100$ task instances.

This design is what enables \bugs{} to be significantly more space-efficient than SWE-bench.
Based on the publicly released images, for SWE-bench's $2294$ task instances, $1.2$ TBs of storage are required to download all Docker images locally.
for SWE-bench Multimodal's $517$ task instances, $1.2$ TBs are required.
The higher per-instance Docker image size for SWE-bench Multimodal is due to how JavaScript dependency management tools (e.g. \texttt{npm}) require more storage compared to equivalent Python infrastructure (e.g. \texttt{pypi}).
\citet{pan_training_2024} states that each image for the $2438$ instances an average of $2.6$GB, totaling 6 TB of storage total.
Such a storage requirement can be a significant barrier for academic practitioners.

On the other hand, with more than $20$x the number of bugs, \bugs{} requires only $125$ Docker images total, corresponding to the number of unique (repository, commit) pairs (in this work, for each repository, we only determine installation and test specifications for one commit).
The $125$ images require a total of $290.54$ GBs.
In summary, compared to SWE-bench's task collection strategy, \bugs{}'s design makes it easier to not only create task instances, but also train on them as well.

\subsection{Case Study: SWE-bench \& \bugs{}}
\label{appx:dataset:case-study}
To better understand the differences between the SWE-bench and \bugs{} collection strategies, we perform \bugs{} collection on the \texttt{pallets/flask} GitHub repository, one of the $12$ test split repositories from the original SWE-bench benchmark.
We review the steps covered in Section~\ref{sec:swesmith:collection} applied to \texttt{pallets/flask} in detail.
First, we defined the installation and testing specifications for the \texttt{pallets/flask} repository at commit \texttt{bc09840}.
Next, we apply the LM modification bug generation strategy to this version of the repository, generating $267$ unique bugs.

We observe several differences.
First, \textit{the \bugs{} collection strategy yields a much higher number of bugs outright.}
From SWE-bench, $11$ task instances are from the \texttt{pallets/flask} repository.
The task instances were originally filtered from $2434$ pull requests (PRs), with $107$ satisfying SWE-bench's filtering criteria of (1) being linked to one or more issues and (2) featuring 1+ new tests.
Out of these $107$, the $11$ ($0.45$\% of $2434$) task instances represent the proportion of PRs that execution environments could be successfully constructed for.
On the other hand, running the function-level rewriting strategy for bug generation originally yielded $402$ candidates, of which $267$ were determined to be valid task instances.

Second, \textit{\bugs{} requires significantly less human effort while only incurring minor costs}.
Collecting the $11$ \texttt{pallets/flask} task instances (steps include scraping PRs, determining repository versions across time, defining version-specific installation/test specifications, running execution-based validation multiple times) took an estimated $38$ hours worth of human labor.
On the contrary, defining installation and testing specifications for the latest commit of \texttt{pallets/flasks} took $10$ minutes. 
The subsequent function-level rewriting strategy for bugs took $23$ minutes to run, incurring a total cost of just \$$2.47$ ($\sim$\$$0.00613$ per instance).
The final execution-based validation step that filters out $402 - 267 = 135$ unqualified bug candidates ran in $14$ minutes.
Since both the bug and problem statement generation strategies are repository agnostic, no additional human intervention is necessary for these steps.
Head to head, per instance for the \texttt{pallets/flask} repository, SWE-bench style collection requires $38 \times 60 / 11 = 207.27$ minutes compared to $0.176$ minutes ($\sim10.6$ seconds) and \$$0.00613$ in API costs using \bugs{}.

Third, \textit{collectively, \bugs{} task instances break a significantly larger proportion of existing tests in a codebase}.
We define ``bug coverage" as the proportion of tests broken by $1$+ instance across all task instances.
For the SWE-bench split of \texttt{pallets/flask}, there are $207$ unique tests across all $11$ instances.
Of these $207$ tests, $15$ are broken by $1$+ instance, corresponding to a bug coverage rate of $7.25$\%.
For the \bugs{} split of \texttt{pallets/flask}, there are $474$ unique tests across $267$ instances.
The larger amount of tests is due to increased test coverage in the \texttt{pallets/flask} repository as of Nov. 28, 2024 (when \bugs{} was collected) compared to June 2023 (when SWE-bench was collected).
Of these $474$ tests, $422$ are broken by $1$+ instance, a bug coverage rate of $89.03$\%.
We attribute the significant difference to a consistent tendency in real world open source software development workflows, that is, the \textit{minority} of tests are introduced to capture existing, errant behavior in the repository.
The significant majority of tests are committed alongside working code, ensuring that already correct behavior is upheld.
Well-maintained repositories will typically not merge commits that cause such tests to fail.
This results in a large number of tests where few to no commits correspond to those tests' failures.

Finally, \textit{\bugs{} does not yield instances appropriate for evaluation}.
The \bugs{} pipeline as presented does not produce hidden tests, a crucial difference that makes SWE-bench more suitable for evaluation.
Consequently, when expert trajectories are generated, the Fail-to-Pass tests are present in the repository at inference time.
Furthermore, our issue generation strategy does not include checks for known problems such as underspecified text descriptions or solution leakage~\citep{chowdhuryintroducing}. 
Simple amendments could make \bugs{} task instances suitable for evaluation, such as deleting Fail-to-Pass test functions or files along with a validation procedure around the ambiguity and leakage of the issue text.
Finally, thorough analyses of how faithful \bugs{} task instances are to real world issues and PRs would be necessary to justify synthetic bugs for evaluation.

\section{Issue Generation}
\label{appx:issue_generation}

We cover the four issue generation strategies we experiment with to determine issue text's effect on how solvable a \bugs{} instance is along with the trajectory's value as a training data point.

\textbf{Generated with LM.}
We prompt an LM with a randomly selected SWE-bench Verified problem statement, the bug patch, list of Fail-to-Pass tests, source code for one Fail-to-Pass test, and the execution logs of running all the Fail-to-Pass tests.
We ask the LM to generate an issue that describes the bug conveyed in the patch in the style of the SWE-bench Verified demonstration.
Figure~\ref{fig:prompt_generate_issue_with_lm} shows the system prompt for this strategy.

\textbf{Fixed issue templates.} We create a set of $7$ pre-defined issue templates, listed in Table~\ref{tab:issue_gen_fixed}.
Each template uses information from the bug patch or Fail-to-Pass tests associated with every task instance.
Given a dataset of task instances, we randomly select one of the templates to use as the problem statement according to the probabilities listed in Table~\ref{tab:issue_gen_fixed}.
The reason we assign the highest likelihood for the prompt that provides all four categories of information (bug type, files changed, functions changed, Fail-to-Pass tests) is to ensure that a higher proportion of task instances are well-specified.

\begin{table}[t]
    \centering
    \begin{tabular}{l|rl}
\toprule
Template & Prob. & Information Provided \\
\midrule
Basic            & $0.05$ & None \\
Files            & $0.1$  & States which file(s) have bug(s). \\
Funcs            & $0.15$ & States which file(s) and func(s) have bug(s). \\
Tests            & $0.1$  & States that some tests are failing. \\
F2P Tests        & $0.1$ & States which tests are failing. \\
Bug Type         & $0.05$ & States failure type. \\
Bug Type + Files & $0.15$ & States failure type and which file(s) have bug(s) \\
Bug Type + Files & $0.15$ & States failure type, which file(s) have bug(s), \\
\quad+ Test      &        & and a random F2P test. \\
Bug Type + Files & $0.15$ & States failure type, which file(s) and func(s) \\
\quad+ Funcs + Test &     & have bug(s), and a random F2P test. \\
\bottomrule
    \end{tabular}
    \caption{
    List of issue text templates we use to generate problem statements.
    Across all templates, four types of information are included --- the files with bugs, functions with bugs, Fail-to-Pass test(s), and the type of bug.
    Templates that offer less information are generally assigned a lower probability.
    }
    \label{tab:issue_gen_fixed}
\end{table}

\textbf{Fail-to-Pass test code and execution logs.} Another approach is showing the source code and test execution logs for a randomly selected Fail-to-Pass test.
This approach is motivated by the lack of reproduction code or expected/actual behavior of code communicated with fixed issue templates.
We show code and execution logs only for a single Fail-to-Pass test; if a task instance has more than one Fail-to-Pass test, we do not disclose remaining tests.

\textbf{Original issue text.}
This strategy works exclusively for some task instances generated using PR Mirroring.
If a PR is successfully mirrored, we use the text from the associated issues as the problem statement, exactly as done in SWE-bench.
Of the $2345$ task instances represented in \bugs{} mirrored from real-world PRs, $708$ or $30.19$\% of these have one or more associated GitHub issue(s) to create a SWE-bench style problem statement.

\begin{example}[System prompt for generating issues with an LM]
\small
  You are a software engineer helping to create a realistic dataset of synthetic GitHub issues. \\
  
  You will be given the following input: \\

  1. Demonstration: A realistic GitHub issue to mimic (included in the \texttt{$<$demonstration$>$} tag). \\
  2. Patch: A git diff output/PR changes that introduces a bug (included in the \texttt{$<$patch$>$} tag). \\
  3. Test output: The output of running the tests after the patch is applied (included in the \texttt{$<$test\_output$>$} tag). \\
  4. Test source code: Source code for one or more tests that failed (included in the \texttt{$<$test\_source\_code$>$} tag). \\

  Output: A realistic GitHub issue for the patch. \\

  Guidelines: \\
  - Mimic the style and structure of the demonstration issues.
    If the demonstration issues are not well structured, your output should also be not well structured.
    If the demonstrations use improper or no markdown, your output should also use improper or no markdown.
    If the demonstrations are short/long, your output should also be short/long (if possible).
    If the demonstrations include human "flavor text" or "fluff", your output should also include human "flavor text" or "fluff".
    Do this even if it conflicts with your default behavior of trying to be extremely concise and helpful. \\
  - DO NOT explain the fix/what caused the bug itself, focus on how to reproduce the issue it introduces \\
  - Do not mention pytest or what exact test failed. Instead, generate a realistic issue. \\
  - If possible, include information about how to reproduce the issue. An ideal reproduction script should raise an error \\
    or print an unexpected output together with the expected output. \\
    However, still include this information in a style very similar to the demonstration issues. \\
\end{example}
\captionof{figure}{
System prompt provided to an LM to generate an issue based off the bug patch and testing information of a task instance along with a demonstration problem statement randomly selected from SWE-bench Verified.
}
\label{fig:prompt_generate_issue_with_lm}

\section{Difficulty Rating}
\label{appx:difficulty}
We train a model that labels a task with one of three difficulty labels: $<15$ minutes (easy), $15$ minutes - $1$ hour (medium), and $1$+ hour (hard).
This model allows us to quantify the difficulty of individual task instances and, in aggregate, the difficulty of entire datasets.

To train this model, we use $1699$ annotations from ~\citet{chowdhuryintroducing}.
In their work towards curating SWE-bench Verified, a subset of $1699$ SWE-bench task instances were labeled with four difficulty levels: $<15$ min, $15$ min - $1$ hr, $1$-$4$ hrs, and $4$+ hrs.
Generally, three annotators were assigned to each instance, and the difficulty annotations were ensembled by taking the majority choice for a sample, or the median if there is no majority.
The distribution of annotated difficulties, from easiest to hardest, is $24.5$\%, $53.5$\%, $19.4$\%, and $2.8$\%.

Because there are very few samples in the $4$+ hr category, we reclassify the $1$-$4$ hr and $4$+ hr instances into a  single $1$+ hr category.
Next, we create corresponding train and test datasets at a $80$/$20$\% split, randomly shuffling the instances while ensuring the train and test distributions do not deviate significantly from the original.
An instance's problem statement and solution patch are provided as input, and one of the three difficulty labels serves as the target output.
We perform LoRA fine-tuning~\citep{hu2021loralowrankadaptationlarge} on a Qwen 2.5 32B Instruct model using the Unsloth~\citep{unsloth} library.
The model achieves an accuracy of $75.3$\% on the test set.
All errant predictions are off by one; in other words, the model never predicted $<15$ min when the label was $1$+ hr, and vise versa.

Using this model, we can grade the difficulty of a \bugs{} instance once the bug patch and corresponding issue text have been created.
To provide a succinct summary of difficulty for a dataset of SWE-bench style task instances, we propose a ``difficulty score" metric.
Each label corresponds to a numeric difficulty score of $1$, $5$, and $9$, from easiest to hardest.
The difficulty score is therefore the average difficulty score across all task instances.

\begin{figure}[t]
    \centering
    \includegraphics[width=\textwidth]{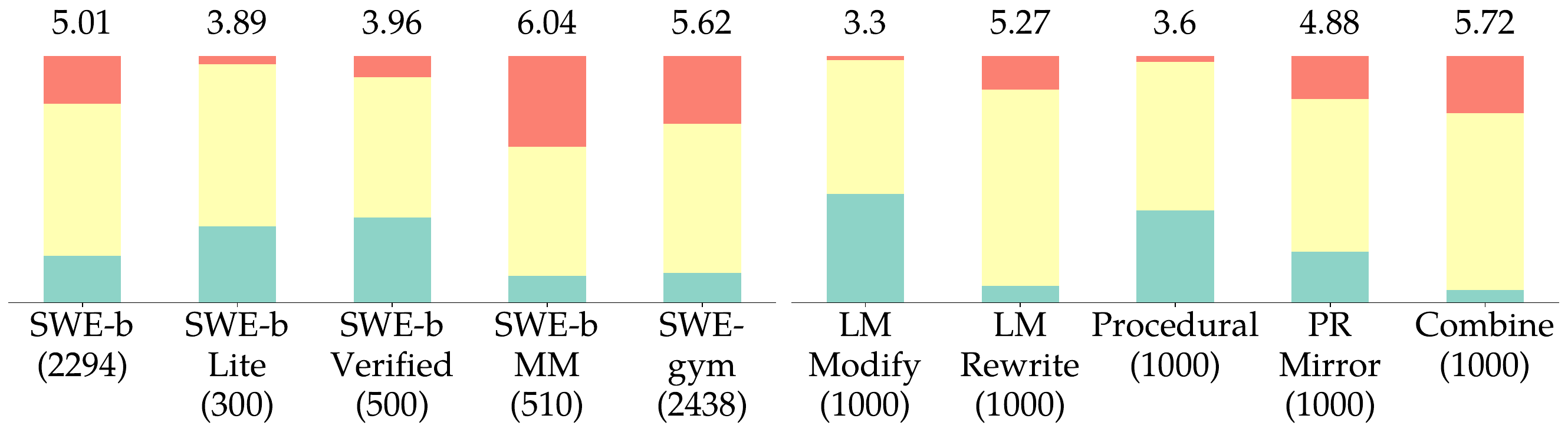}
    \caption{
Distribution of task instance difficulty (\textcolor{green}{\texttt{easy}}/\textcolor{yellow}{\texttt{medium}}/\textcolor{red}{\texttt{hard}}) for existing SWE-bench style datasets (left $5$ bars) and \bugs{} (right $5$ bars), assessed by our difficulty rating model.
The average difficulty score for each dataset is listed above each bar.
For \bugs{}, per bug strategy, we sample $1000$ task instances with LM generated issue text.
    }
    \label{fig:difficulty_scores}
\end{figure}

\begin{table}[b]
    \centering
    \begin{tabular}{l|rrrrr}
    \toprule
        Dataset & \# Instances & Score & \texttt{easy} & \texttt{med} & \texttt{hard} \\
    \midrule
        SWE-bench            & $2294$ & $5.014$ & $438$ & $1408$ & $446$ \\
        \quad Lite           & $300$  & $3.893$ & $93$  & $197$  & $10$ \\
        \quad Verified       & $500$  & $3.960$ & $173$ & $284$  & $43$ \\
        SWE-bench Multimodal & $510$  & $6.036$ & $55$  & $265$  & $186$ \\
        SWE-gym              & $2438$ & $5.625$ & $288$ & $1456$ & $664$ \\
        \quad Lite           & $230$  & $3.890$ & $67$  & $156$  & $4$   \\
    \midrule
        \bugs{} (LM Modify)  & $1000$ & $3.304$ & $441$ & $542$ & $17$ \\
        \bugs{} (LM Rewrite) & $1000$ & $5.272$ & $68$  & $796$ & $136$ \\
        \bugs{} (Procedural) & $1000$ & $3.596$ & $374$ & $603$ & $23$ \\
        \bugs{} (PR Mirror)  & $1000$ & $4.876$ & $206$ & $619$ & $175$ \\
        \bugs{} (Combine)    & $1000$ & $5.720$ & $52$  & $716$ & $232$ \\
    \bottomrule
    \end{tabular}
    \caption{
    The score is averaged over all task instances, where \texttt{easy}/\texttt{med}/\texttt{hard} corresponds to $1$/$5$/$9$.
    For \bugs{}, we sample $1000$ task instances per bug strategy.
    }
    \label{tab:difficulty_scores}
\end{table}

Figure~\ref{fig:difficulty_scores} summarizes our findings for difficulties across different SWE-bench style datasets.
We provide a more thorough rundown of task instances per difficulty level in Table~\ref{tab:difficulty_scores}.
We find that different \bugs{} bug generation methods yield different levels of difficulty.
LM Modify are consistently rated to be easy - from several manual spot checks, we notice that while the prompt for LM Modify provides several examples of types of bugs and does not name specific issues to create, the large majority of bugs created by this strategy are simple variable assignment mistakes (e.g. \texttt{a=a; b=b} is changed to \texttt{a=b; b=a}).
An open-ended prompt like ours does not actually yield high diversity in terms of mistakes created.
Procedural modifications are, as expected, the next easiest, as the types of bugs created by this strategy are finite.
PR Mirrors and LM Rewrites yield much harder tasks, confirmed not only by our bug rating model, but also the lower average resolve rate on these tasks by our expert model (SWE-agent + Claude 3.7 Sonnet).
Finally, aggregating smaller functions together is a simple but effective strategy for creating bugs that are rated as more complex.
This effect aligns with our original expectations; generally, bugs that require editing more functions and files tend to be rated as more difficult.
\bugs{} can be used to create task instances with a range of difficulties.
\section{Experiments}
\label{appx:experiments}
In this section, we provide additional details about the configurations and parameters used to generate trajectories with an expert model and run inference on a fine-tuned model.
We then provide additional ablations and analyses about the \bugs{} dataset and the agents trained on \bugs{}.

\subsection{Training Details}
\label{appx:experiments:train_details}

\textbf{Rejection sampling fine-tuning.} Our fine-tuning setup heavily inherits from ~\citet{pan_training_2024}'s work.
We perform full parameter fine tuning using the \texttt{torchtune}~\citep{torchtune} library, with learning rate \texttt{5e-5}, maximum $3$ epochs, and max context length of $32768$.
Training was carried on Modal~\citep{modal} on $2$-$8$ NVIDIA H100 80G GPUs.
As discussed in Section~\ref{sec:experiments}, the procedure for rejection sampling fine-tuning (RFT) is as follows.
We first generate expert demonstrations/trajectories using SWE-agent and a ``strong" model (e.g. Claude 3.7 Sonnet, GPT 4o) on \bugs{} task instances.
Of these, we then only train a student model on the trajectories corresponding to resolved instances.

\textbf{SWE-agent configuration.}
We use two different configurations, one for generating trajectories with an expert model, and a separate one for running inference on the fine-tuned Qwen, student models.
The configurations are generally quite similar, with minor differences around how LMs' responses are elicited, the parsing mechanism for an LM response, constraints around message sizes, and the system prompt.

We will first review the information common to both configurations.
The prompt template informing an agent of the task's nature and problem statement is included in Figure~\ref{fig:prompt_sweagent_instance}.
This prompt is very similar to the original SWE-agent prompt used in~\citet{yang_swe-agent_2024}.
The prompt templates for showing environment feedback are identical as well.
If there is execution output, the text is simply preceded by \texttt{OBSERVATION: [output]}.
If there is no output (e.g \texttt{rm -r} succeeds silently), then the agent is informed ``Your command ran successfully and did not produce any output".
The agent computer interface (ACI) provided is also identical; SWE-agent provides LM with access to three general tools:
\begin{itemize}[noitemsep]
    \item \texttt{bash}: Execute a bash command in terminal.
    \item \texttt{str\_replace\_editor}: A tool for viewing, creating, and editing files.
    \item \texttt{submit}: A special keyword for the LM to indicate the task is completed or if it is unable to proceed further with the task.
\end{itemize}

\begin{example}[Task Instance Prompt provided to SWE-agent]
$<$uploaded\_files$>$ \\
\{\{working\_dir\}\} \\
$<$/uploaded\_files$>$ \\
I've uploaded a python code repository in the directory \{\{working\_dir\}\}. Consider the following PR description: \\

$<$pr\_description$>$ \\
\{\{problem\_statement\}\} \\
$<$/pr\_description$>$ \\

Can you help me implement the necessary changes to the repository so that the requirements specified in the $<$pr\_description$>$ are met?
I've already taken care of all changes to any of the test files described in the $<$pr\_description$>$. This means you DON'T have to modify the testing logic or any of the tests in any way!
Your task is to make the minimal changes to non-tests files in the \{\{working\_dir\}\} directory to ensure the $<$pr\_description$>$ is satisfied.
Follow these steps to resolve the issue: \\
1. As a first step, it might be a good idea to find and read code relevant to the $<$pr\_description$>$ \\
2. Create a script to reproduce the error and execute it with `python $<$filename.py$>$` using the bash tool, to confirm the error \\
3. Edit the source code of the repo to resolve the issue \\
4. Rerun your reproduce script and confirm that the error is fixed! \\
5. Think about edgecases and make sure your fix handles them as well
Your thinking should be thorough and so it's fine if it's very long.
\end{example}
\captionof{figure}{
A copy of the prompt provided to an LM via SWE-agent informing the LM of the nature of the task, the task description itself, and several tips on how to proceed.
}
\label{fig:prompt_sweagent_instance}

We briefly review the distinctions.
First, tool invocation works differently for expert versus student models.
For the Claude and GPT series models that are used as experts, we use function calling for models to invoke the aforementioned tools.
On the other hand, the student model is asked to generate a response with XML tags to delineate the thought and action.
Therefore, when fine-tuning on expert trajectories, a key processing step is to convert the expert trajectories' function calling format into the XML style response --- fine-tuning \textit{directly} on the expert trajectories does not work.

We note that we use these particular settings because as of the publication of this paper, this tool setting reflects the absolute state-of-the-art performance achieved with an open source agent system (SWE-agent) and any existing LM (Claude 3.7 Sonnet).
It is certainly possible to explore more tool designs and experiment with different formatting calls, as many existing prior works, notably \citet{yang_swe-agent_2024}, have performed.
However, given the focus of our work, we do not bother with repeating such a "hyperparameter sweep" across configurations for the agent system, as this effort is expensive and has already been performed to suggest that the configuration we are using is ideal for expert level performance.

For generating trajectories with expert models, we run with a maximum of $75$ steps and a cost limit of \$$2.00$.
A run terminates automatically when either of these limits are reached or the context window of the expert model is exceeded.
The overwhelming majority of automatic terminations are due to the $75$ maximum steps limit.

For running inference with student models, we run with a maximum of $75$ steps or a cost limit\footnote{We include the cost limit in addition the step limit to provide realistic behavior with respect to handling long context. To calculate a cost value for our model, we use the gpt-4o cost function as of April, 2025.} of \$$2.00$, where the run similarly terminates when either the steps, cost or context window limit is reached.
For the student model, per LM inference call, we truncate the message history to only keep the $5$ most recent tool outputs.
While we occasionally sample trajectories with the expert model set at various temperatures, for the student model, the temperature is fixed at $0.0$.

\subsection{Evaluation Datasets}
\label{appx:experiments:eval_datasets}
\textbf{SWE-bench.} SWE-bench is a widely used benchmark that evaluates AI systems on their ability to resolve GitHub issues~\citep{jimenez_swe-bench_2024}.
Given a codebase along with a description of a bug or feature, the AI system is asked to modify the codebase in such a way that the issue presented in the description is resolved.
SWE-bench consists of $2294$ such task instances, collected from real world pull requests (PRs) and issues in $12$ GitHub repositories that are predominantly Python.
As discussed in Section~\ref{sec:experiments}, the Lite and Verified subsets are curated from the main SWE-bench repository with the goal of making evaluation either more efficent or more reliable.
Since evaluation on the entirety of SWE-bench is fairly costly and does not have as many comparable references, we do not evaluate \texttt{SWE-agent-LM-32B} on the entire SWE-bench test set.

\textbf{SWE-bench Multimodal.} SWE-bench Multimodal applies SWE-bench collection strategy to $12$ additional predominantly JavaScript and TypeScript GitHub repositories, where task instances are associated with issues that have visual asset(s) in them~\citep{yang_swe-bench_2024}.
The evaluation dataset consists of $510$ task instances.
While the original work evaluates vision language models (VLMs) specifically, we do not evaluate \texttt{SWE-agent-LM-32B} which, as it is based on Qwen $2.5$ Coder Instruct, does not have the ability to process images as inputs.

\begin{table}[t]
    \centering
    \begin{minipage}[b]{0.3\textwidth}
        \centering
        \begin{tabular}{ll}
\toprule
\footnotesize jqlang/jq & 9 \\
\footnotesize redis/redis & 12 \\
\scriptsize micropython/micropython & 5 \\
\footnotesize valkey-io/valkey & 4 \\
\footnotesize nlohmann/json & 1 \\
\footnotesize fmtlib/fmt & 11 \\
\midrule
\textbf{C/C++} & 42 \\
\midrule
\scriptsize prometheus/prometheus & 8 \\
\footnotesize caddyserver/caddy & 14 \\
\footnotesize gin-gonic/gin & 8 \\
\footnotesize hashicorp/terraform & 5 \\
\footnotesize gohugoio/hugo & 7 \\
\midrule
\textbf{Go} & 42 \\
\midrule
\footnotesize briannesbitt/carbon & 10 \\
\footnotesize laravel/framework & 13 \\
\scriptsize phpoffice/phpspreadsheet & 10 \\
\scriptsize php-cs-fixer/php-cs-fixer & 10 \\
\midrule
\textbf{PHP} & 43 \\
\bottomrule
        \end{tabular}
    \end{minipage}
    \hfill
    \begin{minipage}[b]{0.3\textwidth}
        \centering
        \begin{tabular}{ll}
\toprule
\footnotesize apache/druid & 5 \\
\footnotesize reactivex/rxjava & 1 \\
\footnotesize apache/lucene & 9 \\
\footnotesize projectlombok/lombok & 17 \\
\footnotesize google/gson & 9 \\
\footnotesize javaparser/javaparser & 2 \\
\midrule
\textbf{Java} & 43 \\
\midrule
\footnotesize babel/babel & 5 \\
\footnotesize mrdoob/three.js & 3 \\
\footnotesize vuejs/core & 5 \\
\footnotesize preactjs/preact & 17 \\
\footnotesize axios/axios & 6 \\
\scriptsize immutable-js/immutable-js & 2 \\
\footnotesize facebook/docusaurus & 5 \\
\midrule
\textbf{JS/TS} & 43 \\
\bottomrule
        \end{tabular}
    \end{minipage}
    \hfill
    \begin{minipage}[b]{0.3\textwidth}
        \centering
        \begin{tabular}{ll}
\toprule
\footnotesize rubocop/rubocop & 16 \\
\footnotesize jekyll/jekyll & 5 \\
\footnotesize faker-ruby/faker & 2 \\
\footnotesize fastlane/fastlane & 7 \\
\footnotesize fluent/fluentd & 12 \\
\footnotesize jordansissel/fpm & 2 \\
\midrule
\textbf{Ruby} & 44 \\
\midrule
\footnotesize tokio-rs/axum & 7 \\
\footnotesize nushell/nushell & 5 \\
\footnotesize sharkdp/bat & 8 \\
\footnotesize burntsushi/ripgrep & 2 \\
\footnotesize uutils/coreutils & 5 \\
\footnotesize tokio-rs/tokio & 9 \\
\footnotesize astral-sh/ruff & 7 \\
\midrule
\textbf{Rust} & 43 \\
\bottomrule
        \end{tabular}
    \end{minipage}
    \caption{
Number of task instances per repository and language in the SWE-bench Multilingual evaluation set.
The entire dataset includes $300$ task instances covering $9$ languages.
    }
    \label{tab:swebml_dataset}
\end{table}

\textbf{SWE-bench Multilingual.} SWE-bench Multilingual is an evaluation dataset consisting of 300 task instances that we introduce with this work.
A single author carried out SWE-bench's collection strategy for $42$ additional GitHub repositories, covering the following $9$ programming languages: JavaScript, TypeScript, C, C++, Go, Java, PHP, Ruby, and Rust. These repositories span a wide range of application domains, including web frameworks, data storage and processing tools, core utilities, and widely used libraries. A brief summary of the dataset is presented in Table~\ref{tab:swebml_dataset}.

Like SWE-bench Verified, we curate the dataset by excluding task instances deemed by a team of three authors to have ambiguous or underspecified issue text. 
Each task instance edits (meaning additions and removals) on average $48$ lines of code.
Similar to SWE-bench and \bugs{}, the median number of Fail-to-Pass tests is one.

We introduce SWE-bench Multilingual to:
\begin{enumerate}
\item Provide a benchmark to evaluate model and agent performance across a variety of programming languages and application domains. Existing agent systems often rely on Python-specific tooling, effectively overfitting to the original SWE-bench~\citep{yang_swe-bench_2024}. Although SWE-bench Multimodal addresses this to some degree, its focus on visual inputs is a confounding factor for text-only evaluation of software engineering capabilities.
\item Remain fully compatible with SWE-bench, so current users can adopt it without changing infrastructure.
\item Keep the dataset small enough to run quickly. While concurrent work like \citet{zan2025multiswebenchmultilingualbenchmarkissue} provides more task instances in multiple languages, we purposely constrain the number of task instances so that the dataset is easy to run quickly.
\end{enumerate}

In \S\ref{appx:experiments:training_analyses}, we briefly discuss how performance by existing state of the art methods for SWE-bench is markedly worse on SWE-bench Multilingual, then offer some clear directions for potential next steps to build better agentic coding models that would involve extending \bugs{}.

\subsection{Trajectory Dataset Breakdown}
\label{appx:experiments:data_breakdown}

\begin{table}[t]
    \centering
    \begin{tabular}{lll|ccc}
\toprule
Purpose & Bug Gen. & Issue Gen. & \# Instances & Temp. & \# Traj. \\
\midrule
\rowcolor{cyan}
\multicolumn{6}{c}{\texttt{claude-3-7-sonnet-20250219}} \\
\midrule
Ablation       & LM (Modify)  & LM       & 1000 & 0 & 605 \\
(Bug Type)     & LM (Rewrite) & LM       & 1000 & 0 & 507 \\
               & Procedural   & LM       & 1000 & 0 & 745 \\
               & PR Mirrors   & LM       & 1000 & 0 & 557 \\
\midrule
Ablation       & PR Mirrors   & Fixed    & 600  & 0 & 259 \\
(Issue Type)   & PR Mirrors   & F2P Test & 600  & 0 & 390 \\ 
               & PR Mirrors   & Original & 600  & 0 & 328 \\ 
               & PR Mirrors   & LM       & 600  & 0 & 319 \\
\midrule
Ablation       & Procedural   & LM       & 1000 & 0 & 721 \\
(Repositories) & Procedural   & LM       & 1000 & 0 & 709 \\
               & Procedural   & LM       & 1000 & 0 & 723 \\
               & Procedural   & LM       & 1000 & 0 & 707 \\
\midrule
Final Dataset  & LM (Rewrite) & LM       & 3574 & 0 & 1003 \\
Curation       & PR Mirrors   & LM       & 1049 & 0 & 349 \\
\midrule
\rowcolor{cyan}
\multicolumn{6}{c}{\texttt{claude-3-5-sonnet-20250219}} \\
\midrule
Compare with prior work & All & LM & 800 & 0 & 535 \\
\midrule
\rowcolor{cyan}
\multicolumn{6}{c}{\texttt{gpt-4o-2024-08-06}} \\
\midrule
Compare with prior work & All & LM & 200 & 0 & 89 \\
\bottomrule
    \end{tabular}
    \caption{
    Breakdown of trajectories sampled from \bugs{}.
    Trajectories were generated from subsets of \bugs{} that were either for the purpose of ablations or performance.
    All trajectories were generated with a maximum of $75$ steps and a \$$2$ cost limit.
    }
    \label{tab:trajectories_breakdown}
\end{table}

\begin{table}[t]
\centering
\begin{minipage}[t]{0.38\textwidth}
    \centering
    \begin{tabular}{l|c}
\toprule
Bug Type & Count \\
\midrule
Combine (File)   & 123  \\
Combine (Module) & 7    \\
LM (Modify)      & 11   \\
LM (Rewrite)     & 1532 \\
Procedural       & 1495 \\
PR Mirror        & 1848 \\
\bottomrule
    \end{tabular}
    \caption{
Bug types represented in final training dataset.
    }
    \label{tab:sft_final_bug_types}
\end{minipage}
\hfill
\begin{minipage}[t]{0.59\textwidth}
    \centering
    \begin{tabular}{lc|lc}
\toprule
Repository & Count & Repository & Count \\
\midrule
\scriptsize getmoto/moto & $378$ & \scriptsize sqlfluff/sqlfluff & $122$ \\
\scriptsize pandas-dev/pandas & $320$ & \scriptsize pylint-dev/astroid & $110$ \\
\scriptsize conan-io/conan & $243$ & \scriptsize pydicom/pydicom & $103$ \\
\scriptsize pydantic/pydantic & $209$ & \scriptsize tobymao/sqlglot & $101$ \\
\scriptsize iterative/dvc & $181$ & \scriptsize pygments/pygments & $99$ \\
\scriptsize dask/dask & $139$ & \scriptsize scanny/python-pptx & $98$ \\
\bottomrule
    \end{tabular}
    \caption{
Top ten repositories by number of trajectories represented in final dataset for main result.
    }
    \label{tab:sft_final_repos}
\end{minipage}
\end{table}

We provide a thorough review of the dataset of SWE-agent trajectories released with this work in Table~\ref{tab:trajectories_breakdown}.
The majority are generated with \texttt{claude-3-7-sonnet-20250219}.
To compare with prior work, a minority were generated with \texttt{claude-3-5-sonnet-20240620} and \texttt{gpt-4o-2024-08-06}.
As mentioned in Section~\ref{sec:results}, to guard against the easy data bias phenomenon, we impose a per-instance cap of $3$, meaning for any task instance, we include at most $3$ trajectories successfully resolving that task instance in our fine-tuning dataset.
From the pool of trajectories reflected in Table~\ref{tab:trajectories_breakdown}, we curate a set of $5000$ trajectories that we then use to train \texttt{SWE-agent-LM-32B}.

\begin{figure}[t]
    \centering
    \includegraphics[width=0.85\linewidth]{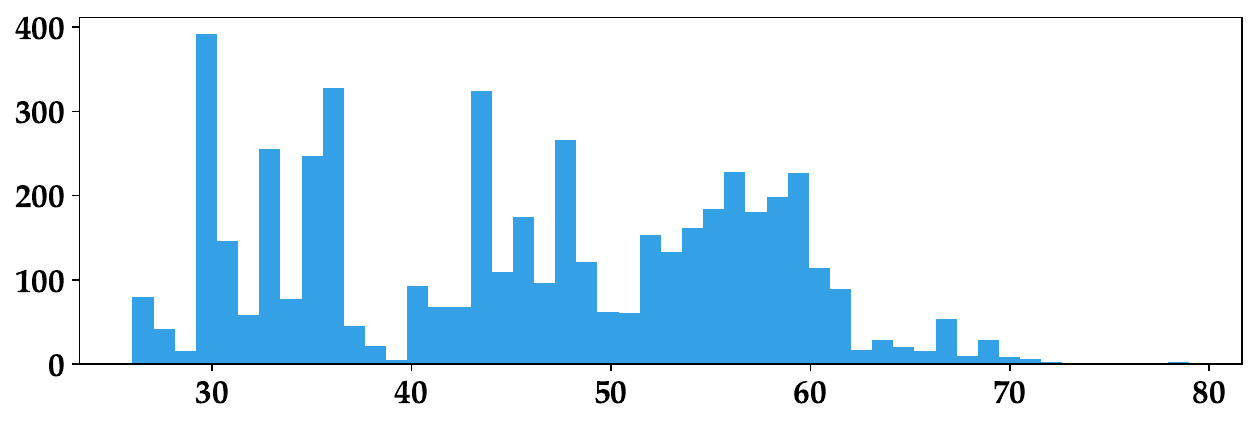}
    \caption{Distribution of number of turns for trajectories represented in the final dataset.}
    \label{fig:num_turns_dist}
\end{figure}

Tables~\ref{tab:sft_final_bug_types} and~\ref{tab:sft_final_repos} show what repositories and bug types are represented in the final training dataset.
In total, $123$ repositories are represented, with at least $10$ trajectories from $91$ repositories.
Trajectories are on average $58$ turns long, meaning an LM typically takes $29$ actions for a given demonstration trajectory.
We visualize this distribution in Figure~\ref{fig:num_turns_dist}.

\subsection{Training Analyses}
\label{appx:experiments:training_analyses}
We provide additional experiments and discussions around training \texttt{SWE-agent-LM-32B}.

\begin{figure}[t]
\centering
    \begin{minipage}[t]{0.49\textwidth}
        \includegraphics[width=\linewidth]{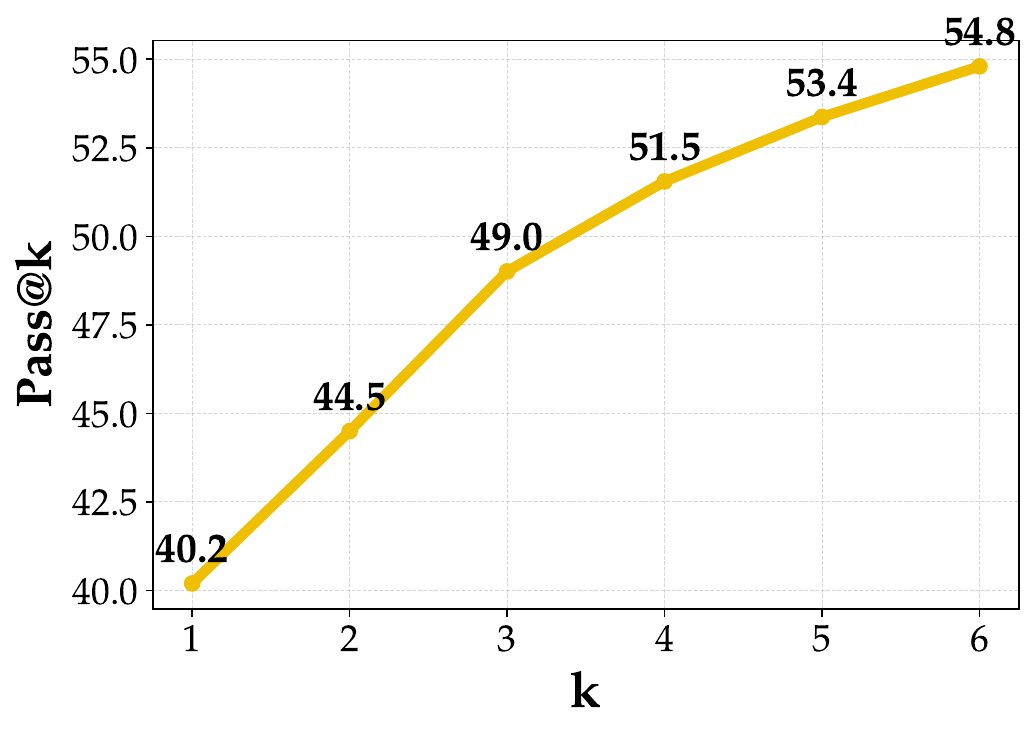}
        \caption{
        \texttt{SWE-agent-LM-32B} Pass@k curve on SWE-bench Verified.
        We observe higher \% resolved when considering
        more runs.
        }
        \label{fig:pass_at_k}
    \end{minipage}
    \hfill
    \begin{minipage}[t]{0.49\textwidth}
        \includegraphics[width=\linewidth]{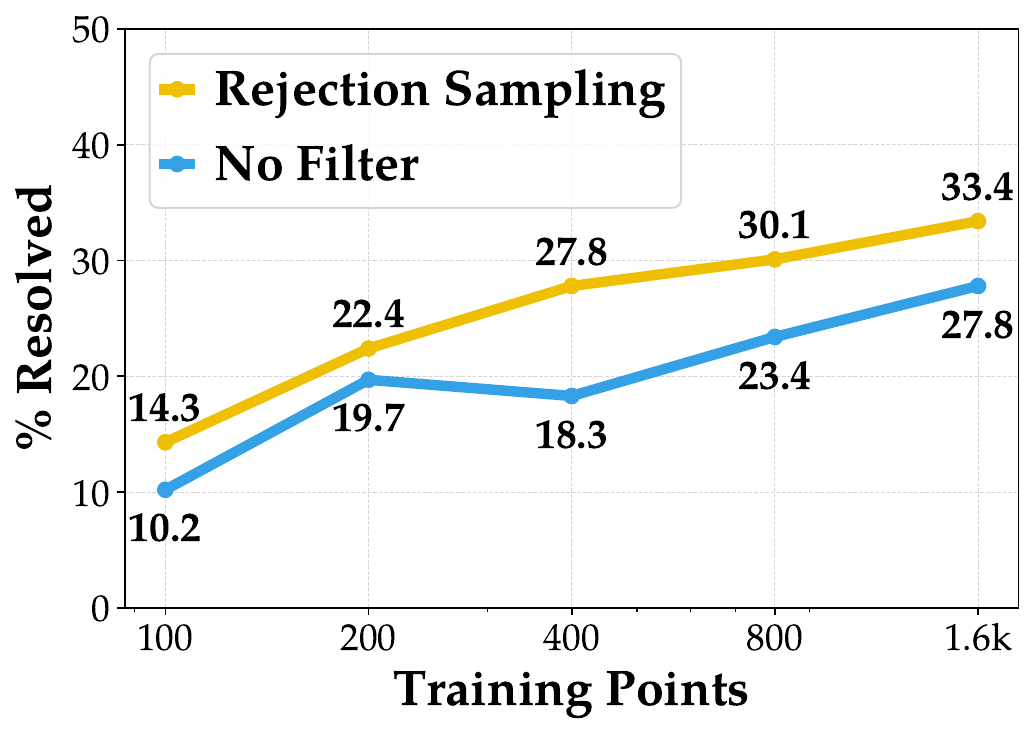}
        \caption{
        Rejection sampling fine-tuning
        leads to better performance than random sampling
        of trajectories for training.
        }
        \label{fig:sft_vs_rft}
    \end{minipage}
\end{figure}

\textbf{Pass@k trend line.} To calculate the Pass@1 score discussed in our main result, we ran SWE-agent with \texttt{SWE-agent-LM-32B} six times.
In Figure~\ref{fig:pass_at_k}, we observe increasing performance at higher values of \texttt{k}, a phenomenon that reflects observations in prior works across LMs for software engineering, code generation, web navigation, and theorem proving.
While we do not explore work around inference time scaling and training a separate verifier model to select the best solution candidate generated by multiple roll-outs, as done in~\citet{pan_training_2024} and ~\citet{jain2025r2e-gym}, \texttt{SWE-agent-LM-32B} is fully compatible with the generate-then-select pipelines explored by such works.
Given its strong Pass@1 performance, \texttt{SWE-agent-LM-32B} would likely be quite competitive for Best@k results as well.
As mentioned before, all trajectories generated in the course of \bugs{} have been released publicly, which the community might find useful for training better verifiers.

\textbf{Rejection sampling fine-tuning ablation.} To confirm that rejection sampling fine-tuning leads to better performance on the downstream task, we compare against a setting where we randomly sample \texttt{n} training points with no filtering criteria, at \texttt{n = [100, 200, 400, 800, 1600]} and fine-tune the same student model (Qwen $2.5$ Coder Instruct $32$B.
We then run SWE-agent with each student model on the SWE-bench Verified dataset three times, with the ``\% Resolved" corresponding to the Pass@1 score.
We show results in Figure~\ref{fig:sft_vs_rft}, which confirms that fine-tuning only on trajectories corresponding to successfully resolved tasks is better than randomly sampling trajectories.

\textbf{SWE-bench Multilingual performance.} To assess how well \texttt{SWE-agent-LM-32B} and existing models generalize to non-Python coding domains, we evaluate the performance of our model, Qwen $2.5$ Coder Instruct 32B, and Claude $3.7$ Sonnet with SWE-agent on our new dataset, which we introduced in Section~\ref{appx:experiments:eval_datasets}.
Out of $300$ task instances, we found that Claude $3.7$ Sonnet achieved a $43$\% Pass@1 resolve rate, which is significantly better than \texttt{SWE-agent-LM-32B} ($8.4$\%) and Qwen $2.5$ Coder Instruct ($6.5$\%).
\texttt{SWE-agent-LM-32B} does not demonstrate a significant improvement over the baseline model.
Through several spot checks of different trajectories, we came to a working hypothesis that while the rejection sampling fine-tuning process had improved its ability to carry out multi-turn interactions in this task setting, there were instances where code edits reflected syntax closer to Python despite code and files viewed in previous steps clearly not being written in Python.

While the result for \texttt{SWE-agent-LM-32B} SWE-bench Multilingual is clearly subpar, we are excited by such a finding, as it motivates future work on top of \bugs{}.
To elaborate, we expect that the path to open agent coding models capable of generalizing to many repositories and languages will be paved by more data and better training techniques, both of which \bugs{} is very capable of facilitating.
First, regarding data, although we wrote \bugs{} to be Python centric, the collection methodology and bug generation techniques (especially LM based methods) should be readily transferable to other repositories.
Second, the negative result on SWE-bench Multilingual provides a clear impetus for exploring whether better training techniques could lead to models that are trained on one code domain (e.g., Python), but can generalize to many languages and repositories.

\subsection{Agent Behavioral Studies}
\label{appendix:experiments:agentbehavior}
\subsubsection{Turn counts and cost}
\label{appendix:experiments:agentbehavior:turncounts}
\begin{figure}[t]
    \begin{minipage}[t]{0.49\textwidth}
        \includegraphics[scale=0.5]{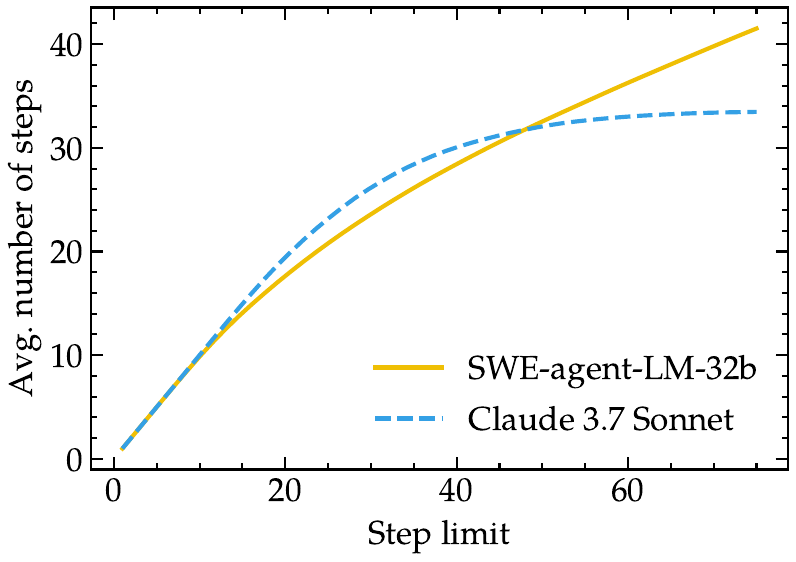}
    \caption{The average step count depends strongly on the prescribed step limit.}
    \label{fig:step_cutoff_vs_avg_steps}
    \end{minipage}
    \hfill
    \begin{minipage}[t]{0.49\textwidth}
        \includegraphics[scale=0.5]{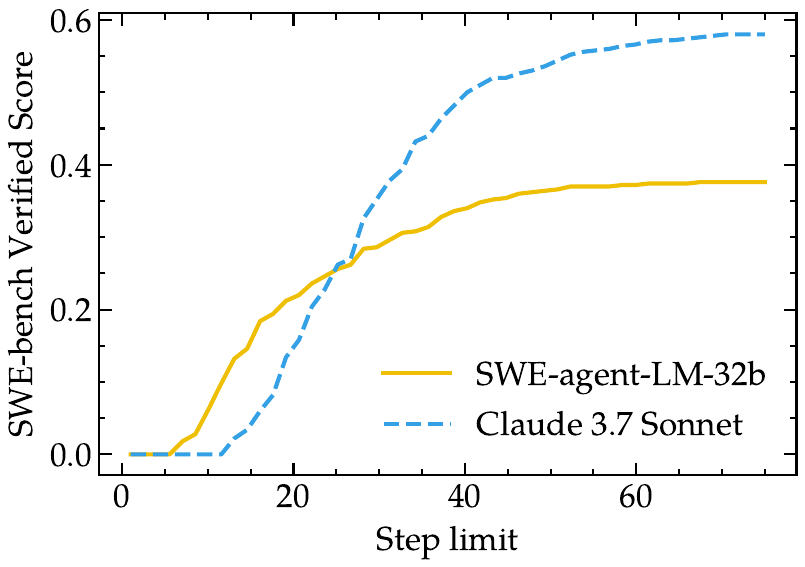}
    \caption{Number of successful instances submitted before a given step limit.}
    \end{minipage}
\end{figure}

While agents are frequently quoted with a singular cost-per-instance number, this can be very misleading in the case of SWE-agent-LM-32B.
Because most of the failed instances fail due to termination by the cost or turn count limit, the average cost and turn counts depend strongly on these limits (see Fig.~\ref{fig:step_cutoff_vs_avg_steps}).

We can also chart the number of resolved instances vs step limits.
To avoid reevaluating the agent with multiple step limits, we use one run with step limit 75 and then assume that a successful agent run that terminates after step $n$ would have failed when restricted by a limit smaller than $n$.
This chart corroborates the point made in section~\ref{sec:experiments}: SWE-agent-LM-32B has a higher resolution rate for very low step limits. 

\subsubsection{Analysis of agent action space}
\textbf{Reduction to \emph{base commands.}} In addition to the dedicated tools provided to the agent as part of the agent computer interface (Section~\ref{appx:experiments:train_details}), the agent can execute arbitrary bash commands.
This makes quantitative analyses of the agent action space challenging.
For example, the agent might issue commands like \texttt{PYTHONPATH=/testbed/repo cd /testbed/repo \&\& python3 reproduce.py}.
We have found the following procedure to determine a \emph{base command} effective to meaningfully describe the action:
\begin{enumerate}
    \item Strip any environment variable manipulation from the beginning of the command.
    \item When multiple commands are chained with \texttt{\&\&} or semicolons, only consider the last command.
    \item Remove all arguments. Because some commands have subcommands (e.g., \texttt{git checkout}), we apply several basic heuristics to determine whether to keep the first or the first two words.
\end{enumerate}

\textbf{Repetitive actions.} 
We determine the longest repetitive sequence of actions by determining the longest sequence of identical base commands within the agent actions.
Note that this means that e.g., \texttt{str\_replace\_editor view} actions that target different files are considered to be repetitive actions as far as this analysis is concerned. 

\subsubsection{Failure mode analysis}

\begin{figure}[t]
    \centering
    \includegraphics[width=\textwidth]{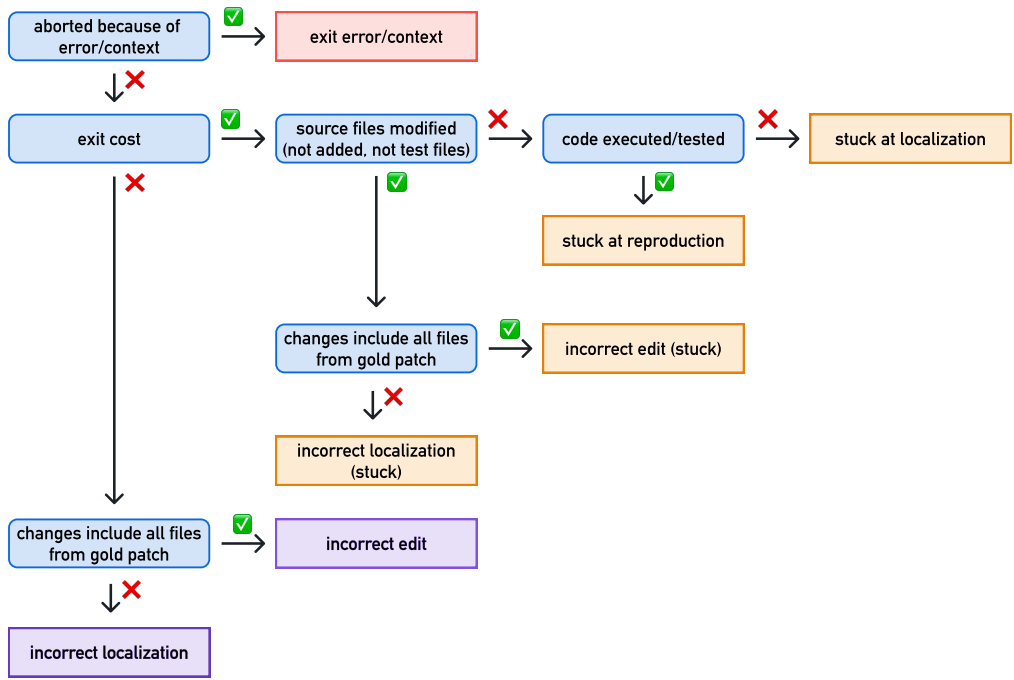}
    \caption{Categorizing failure modes}
    \label{fig:failure_mode_analysis}
\end{figure}

Categorizing the failure mode proceeds as shown in Figure~\ref{fig:failure_mode_analysis}:
\begin{enumerate}
\item \textbf{Error conditions:} If the agent terminates due to an error (environment errors, inability of the LM to correctly format its messages, etc.) or because it exceeded its maximum context window, we return the \textbf{error} or \textbf{context} category.
\item \textbf{Early termination:} If the agent was terminated because of a step or cost limit, we return one of the \textbf{stuck …} subcategories. Note that the SWE-agent still attempts to extract a submission (list of changes/patch).
We determine the subcategory based on which part of the workflow agentic loop was terminated:
\begin{enumerate}
\item If no source (i.e., non-test) file was modified\footnote{We exclude added files because solving SWE-bench instances always requires \emph{changes} to existing files.} and no attempt at testing was made, we return \textbf{stuck at localization}. If test commands were run (i.e., \texttt{python}, \texttt{pytest}, \dots, or similar commands), we return \textbf{stuck at reproduction}.
\item If source files \emph{were} modified, we check whether the changes include changes to all source files that are modified in the gold patch. If not, we return \textbf{incorrect localization (stuck)}, else \textbf{incorrect edit (stuck)}.
\end{enumerate}
\item \textbf{Successful submission:} If the agent terminated and submitted a solution naturally, we return \textbf{incorrect localization} or \textbf{incorrect edit}, depending on whether the changes from the submitted patch included changes to all files from the SWE-bench gold patch.
\end{enumerate}

\subsubsection{Mitigating repetitive actions}
As described in section~\ref{sec:results}, \texttt{SWE-agent-LM-32B} frequently shows highly repetitive actions for unresolved instances.
In light of this, it seems promising to investigate whether agent scaffolding interventions can be used to mitigate the problem and increase the success rates.

\begin{figure}[t]
\centering
   \includegraphics[width=0.49\textwidth]{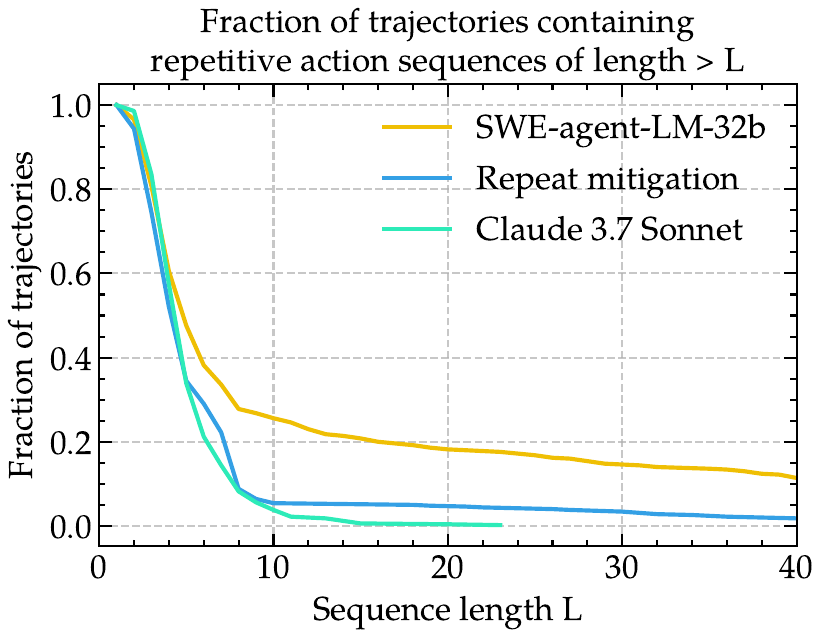}
   \caption{Scaffold interventions can drastically reduce the number of repetitive actions.}
   \label{fig:app:repeat_mitigation}
\end{figure}

We make the following modification to the agent scaffold:
\begin{itemize}
    \item We add warning messages to the observation (command output) if a base command is repeated four (\texttt{str\_replace\_editor view}) or six (any other base command) times. The warning message advises to try different commands, and in particular suggest to locate relevant context using \texttt{find} or \texttt{grep}. 
    \item If the warning messages do not break the string of repetitive base commands and the repetition length reaches 6 (\texttt{str\_replace\_editor view}) or 8 (any other base command), every following action is resampled up to 10 times, stopping at the first base command that is distinct from the previous ones. 
    To further increase the likelihood of breaking the cycle, we inject assistant messages or raise the temperature if the repetition length reaches 7 or 9.
\end{itemize}
This effectively reduces the number of repetitive actions (see Fig.~\ref{fig:app:repeat_mitigation}).
However, the overall number of resolved instances drops slightly to 192 ($38.4\%$).
Variations of the above strategies yield similar outcomes: while repetition is suppressed, success rates do not improve substantially.
This may suggest that repetitive actions are better understood as \emph{symptoms} of the model’s difficulty in solving an instance (such as when the instance is out-of-distribution or particularly challenging) rather than constituting intrinsic failure modes.
\section{Miscellaneous}
\label{appx:misc}

\textbf{Teaser figure description.} We briefly describe how the left hand graph of Figure~\ref{fig:scaling_teaser}, which depicts scaling of task instance collection for the \bugs{} vs. SWE-bench, was created.
For \bugs{}, we simply collected the number of task instances for each repository.
For SWE-bench, we ran the SWE-bench task instance candidate collection script on all $128$ repositories, which first crawls all PRs from a given repository.
Then, each PR that edits at least one or more Python files and changes at least one or more testing related files is converted into a candidate task instance.
Finally, based on the average task instance yield rate reported in~\citet{jimenez_swe-bench_2024}, we estimate the number of viable task instances to be $20$\% of the candidates.
We then determine the number of task instances for \texttt{n} repositories at intervals of $5$ repositories ranging from $5$ to $250$, where the repositories are sorted by number of stars.
In other words, the first five repositories we account for in the figure are the five with the fewest number of stars out of the $128$ repositories used.

\textbf{Extended related works.}
We discuss additional related works briefly, primarily about similar work towards synthesizing trajectories for training LM agents, but for the domain of web tasks.
To improve the interactive capabilities of open source LMs~\citep{chen_fireact_2023}, prior works have also explored trajectory generation techniques for web benchmarks and settings~\citep{xie_osworld_2024,yao_webshop_2023,zhou_webarena_2024}.
For web navigation, existing strategies rely on (1) performing random walks which are then labeled retroactively with instructions~\citep{xiang_language_2023,murty_nnetscape_2024}, (2) using online web tutorials as a source of indirect supervision for generating synthetic trajectories~\citep{ou_synatra_2024}, or (3) collecting human demonstrations~\citep{shen2024scribeagentspecializedwebagents,xu_agenttrek_2024}.
These procedures do not translate well to the software engineering setting; random sequences of command line interactions usually do not achieve meaningful effects on a codebase.
Our cursory efforts around replaying trajectories synthesized from online code edit sequences (e.g. GitHub commit histories) were unsuccessful due to the limited information available, which primarily capture file-level changes without reflecting the underlying skills, decision-making, or the broader context of a software development process.

Our exploration of using SWE-agent to automatically determine installation and testing specifications for a repository is heavily influenced by two research directions - automatic execution environment construction using LMs~\citep{bogin2024superevaluatingagentssetting,eliseeva2025envbenchbenchmarkautomatedenvironment,vergopoulos2025automatedbenchmarkgenerationrepositorylevel}, and generating unit tests using LMs~\citep{mündler2025swtbenchtestingvalidatingrealworld}.
Although relatively much less than SWE-bench style collection, \bugs{} still requires minimal amounts of human labor (around $8$ minutes total per repository).
As we expand \bugs{} to more repositories and languages, we are continuing to consider how to completely automate the environment construction process end to end.

\end{document}